\documentclass[apj]{emulateapj}
\slugcomment{Received 2013 August 7; accepted 2014 February 3; published 2014 March ?}

\usepackage{amsmath}
\usepackage{color}
\usepackage[bookmarks,bookmarksnumbered,colorlinks=true, citecolor=blue, linkcolor=black]{hyperref}

\newcommand{\msun}{\,\rm M_\odot}

\newcommand{\f}{\frac} 

\newcommand{\mvir}{M_{\rm halo}}

\newcommand{\kms}{\ifmmode \,{\rm km\,s^{-1}}\else km$\,$s$^{-1}$\fi}

\newcommand{\vR}{\ifmmode v_R \else $v_R$\fi}
\newcommand{\vtheta}{\ifmmode v_\theta \else $v_\theta$\fi}
\newcommand{\vz}{\ifmmode v_z \else $v_z$\fi}
\newcommand{\vpeak}{\ifmmode v_{\rm peak} \else $v_{\rm peak}$\fi}
\newcommand{\vesc}{\ifmmode v_{\rm esc} \else $v_{\rm esc}$\fi}
\newcommand{\fv}{\ifmmode f(v) \else $f(v)$\fi}
\newcommand{\rhodm}{\ifmmode \rho_\chi \else $\rho_\chi$\fi}
\newcommand{\vmin}{\ifmmode v_{\rm min} \else $v_{\rm min}$\fi}
\newcommand{\gvmin}{\ifmmode g(v_{\rm min}) \else $g(v_{\rm min})$\fi}
\newcommand{\Asym}{{\rm Asym}}

\lefthead{Pillepich et al.}
\righthead{}

\begin{document}

\title{The Distribution of Dark Matter in the Milky Way's Disk}

\author{Annalisa Pillepich\altaffilmark{1,2}, Michael Kuhlen\altaffilmark{1,3}, Javiera Guedes\altaffilmark{4}, and Piero Madau\altaffilmark{1}}

\affil{$^1$Department of Astronomy \& Astrophysics, University of California Santa Cruz, 1156 High St., Santa Cruz, CA 95064 \\
	$^2$Harvard--Smithsonian Center for Astrophysics, 60 Garden Street, Cambridge, MA 02138\\
       $^3$Theoretical Astrophysics Center, University of California Berkeley, Hearst Field Annex, Berkeley, CA 94720 \\
       $^4$ETH Zurich, Institute for Astronomy, Wolfgang-Pauli-Strasse 27, Zurich 8049, Switzerland}

\email{apillepich@cfa.harvard.edu}

\begin{abstract}
We present an analysis of the effects of dissipational baryonic physics on the local dark matter (DM) distribution at the location of the Sun, with an emphasis on the consequences for direct detection experiments. Our work is based on a comparative analysis of two cosmological simulations with identical initial conditions of a Milky Way halo, one of which (Eris) is a full hydrodynamic simulation and the other (ErisDark) is a DM-only one. We find that two distinct processes lead in Eris to a 30\% enhancement of DM in the disk plane at the location of the Sun: the accretion and disruption of satellites resulting in a DM component with net angular momentum and the contraction of baryons pulling DM into the disk plane without forcing it to co-rotate. Owing to its particularly quiescent merger history for dark halos of Milky Way mass, the co-rotating dark disk in Eris is less massive than what has been suggested by previous work, contributing only 9\% of the local DM density. Yet, since the simulation results in a realistic Milky Way analog galaxy, its DM halo provides a plausible alternative to the Maxwellian standard halo model (SHM) commonly used in direct detection analyses. The speed distribution in Eris is broadened and shifted to higher speeds compared to its DM-only twin simulation ErisDark. At high speeds $f(v)$ falls more steeply in Eris than in ErisDark or the SHM, easing the tension between recent results from the CDMS-II and XENON100 experiments. The non-Maxwellian aspects of $f(v)$ are still present, but much less pronounced in Eris than in DM-only runs. The weak dark disk increases the time-averaged scattering rate by only a few percent at low recoil energies. On the high velocity tail, however, the increase in typical speeds due to baryonic contraction results in strongly enhanced mean scattering rates compared to ErisDark, although they are still suppressed compared to the SHM. Similar trends are seen regarding the amplitude of the annual modulation, while the modulated fraction is increased compared to the SHM and decreased compared to ErisDark.
\end{abstract}

 
\section{Introduction}

The direct detection of dark matter (DM) is one of the most exciting and frontier pursuits of contemporary physics. Direct detection experiments attempt to measure the weak nuclear recoils produced in rare scatterings of DM particles off target nuclei in shielded underground terrestrial detectors \citep{goodman_detectability_1985,gaitskell_direct_2004}. After many years of steady progress in enlarging target masses, improving detector sensitivities, and lowering energy thresholds, but concomitant lack of detections and only ever more stringent exclusion limits, the field may now at last be at the cusp of success. In addition to the long standing detection claim by the DAMA collaboration \citep{bernabei_search_2000,bernabei_new_2010}, a number of additional experiments have in recent years reported signals that may be interpreted as DM scattering events.

Specifically, the CoGeNT collaboration has reported a statistically significant excess of events over their well characterized radioactive background \citep{aalseth_cogent:_2012}, together with an annual modulation signal at somewhat lower significance \citep{aalseth_search_2011}. Similarly, the CRESST-II experiment has reported 67 events in their signal acceptance region \citep{angloher_results_2012}, which cannot be accounted for by known backgrounds at a statistical significance of more than $4 \sigma$, yet match the expectation of DM scattering events. Finally, a recent analysis from the CDMS II collaboration of data obtained with their silicon detectors found three DM candidate events with a total expected background of 0.7 events \citep{cdms_collaboration_dark_2013}. Taking into account the energies of the three events, the CDMS II Si data prefer a DM scattering interpretation over a known-background-only scenario at 99.81\% probability, so slightly more than $3 \sigma$.

Despite these exciting developments, the case for a discovery of a DM particle is not yet closed, for two principal reasons. For one, the regions of parameter space (mass of DM particle $m_\chi$ and (spin-independent) scattering cross section $\sigma_{\rm SI}$) preferred by the tentative detections don't all agree with each other. They do generally favor a light DM particle ($m_\chi \lesssim 10$ GeV) with $\sigma_{\rm SI}$ around $10^{-41}\text{--}10^{-40} \, {\rm cm}^2$, but the published $2 \sigma$ confidence intervals don't all overlap \citep[for a recent summary, see e.g. Fig.4 of][]{cdms_collaboration_dark_2013}. Secondly, the preferred parameters are nominally ruled out by the non-detections in XENON100 \citep{aprile_dark_2012} and the CDMS II Germanium detectors \citep{cdms_ii_collaboration_dark_2010,ahmed_results_2011}.\footnote{Note, however, that alternative interpretations of the XENON100 and CDMS II (Ge) data exist: a re-analysis of the low energy Ge events in CDMS II by \citet{collar_maximum_2012} finds strong evidence ($5.6 \sigma$) for a population of nuclear recoil events, and \citet{hooper_revisiting_2013} makes the point that the two nuclear recoil candidate events in the XENON100 data are more easily explained as DM scattering events rather than background leakage.}

All direct detection analyses must make an assumption about the local phase-space distribution of the DM particles incident on Earth. The most commonly used model is the so-called Standard Halo Model (SHM), in which the local DM density is taken to be $\rho_0 = 0.3$ GeV cm$^{-3}$ \citep[consistent with the most recent observational constraints,][]{garbari_limits_2011, garbari_limits_2012, zhang_segue_2013, bovy_rix_2013} and the halo rest-frame speed distribution \fv\ is assumed to be a Maxwellian with a peak (most probable) speed of 220 \kms\ and a cutoff at the Galactic escape speed of 550 \kms. A consistent interpretation of the various detection claims and exclusion limits is complicated by the fact that the experiments (with different target nuclei and energy thresholds) are sensitive to different speed ranges, and thus depend on the assumed \fv\ in different ways. Indeed, departures from the Maxwellian assumption may allow some of the conflicting detection claims to be reconciled \citep{frandsen_resolving_2012,kelso_toward_2012,mao_connecting_2013}. Although it is possible to compare results from multiple experiments in a way that is independent of astrophysical assumptions \citep[see e.g.][]{fox_integrating_2011,frandsen_resolving_2012}, this technique only applies over the limited range of recoil energies for which the experiments probe the same region of \fv.

Numerical galaxy formation simulations can provide guidance for the expected local DM density and velocity distribution, and their spatial and halo-to-halo variance. Although ultra-high resolution DM-only cosmological simulations like Via Lactea II \citep{diemand_clumps_2008} and Aquarius \citep{springel_aquarius_2008} predict that the Milky Way's halo should be filled with a large number of dense self-bound subhalos, the simulations tend to find that the central regions near the location of the Sun remain quite smooth \citep{zemp_graininess_2009}, owing to the strong tidal forces that tend to disrupt subhalos. 
The relics of such disrupted subhalos are predicted in turn to traverse the Solar neighborhood under the form of thousands of DM streams \citep{vogelsberger_phase-space_2009, fantin_finestructure_2011}, however their superposition is also expected to be smooth.
It thus appears unlikely that the Earth lies inside a significant over- or under-density with respect to mean density at 8 kpc \citep{kamionkowski_galactic_2008,kuhlen_dark_2010,kamionkowski_galactic_2010}, and the SHM is acceptable in this regard.

The situation is quite different for the velocity distribution. Here numerical simulations have pointed out a number of departures from the Maxwellian shape assumed in the SHM. The speed distributions averaged in a spherical shell at 8 kpc in DM-only simulations typically shows a pronounced deficit near the peak and an excess on the high speed tail, before again falling below the Maxwellian at the highest speeds \citep{hansen_universal_2006,vogelsberger_phase-space_2009,kuhlen_dark_2010}. The high speed excess arises in part from a ``debris flow'' \citep{kuhlen_direct_2012}, and it reflects the incompletely phase-mixed nature of the DM halo. The simulated \fv\ is much better described by a Tsallis distribution \citep{vergados_impact_2008}, a modified Gaussian distribution \citep{fairbairn_spin-independent_2009}, or the empirical fitting function proposed by \citet{mao_halo--halo_2013}. Furthermore the shape of \fv\ depends on the location within the halo relative to its scale radius and exhibits considerable scatter between halos \citep{mao_halo--halo_2013}. In addition to these global non-Maxwellian features, velocity space substructure can give rise to spatial variations in \fv, with individual subhalos or tidal streams producing spikes at discrete speeds \citep{kuhlen_dark_2010}. DM associated with the Sagittarius tidal stream is an example of a known velocity substructure in our Galaxy that is likely to have a non-negligible influence on DM detection experiments \citep{purcell_dark_2012}.

Most of the above results are based on DM-only simulations, which neglect the effects of baryonic physics in order to achieve extremely high spatial and mass resolution. In recent years, however, increasing computational resources and advances in the treatment of baryonic physics have made it possible to follow the formation of disk galaxies like our Milky Way in cosmological simulations that include dissipational gas physics \citep[e.g.][]{governato_bulgeless_2010,agertz_formation_2011,guedes_forming_2011}. Because baryons are able to radiate energy and condense in the centers of halos, they have the potential to modify the structure of their hosting dark matter halos, and may thereby alter the expectations for direct detection experiments. In particular, adiabatic contraction \citep{blumenthal_contraction_1986,gnedin_response_2004} may drag DM toward the halo center and thus increase the local DM density. On the other hand, violent energetic feedback processes might result in the removal of DM from halo centers and the formation of a DM core \citep{read_mass_2005,mashchenko_stellar_2008,governato_bulgeless_2010,pontzen_how_2012}. Baryonic physics may even result in an offset between the point of maximum DM density and the dynamical center of the Galaxy \citep{kuhlen_off-center_2013}.

Regarding baryonic modifications of the local DM velocity structure, the effect that has received the most attention is the possibility of the creation of a so-called ``dark disk'' \citep{lake_darkdisk_1989, read_thin_2008, read_dark_2009, purcell_dark_2009, ling_dark_2010} -- a flattened component of the DM halo, nearly co-rotating with the stellar disk, that is thought to be formed by the tidal disruption of accreted satellites dragged into the disk plane by dynamical friction. The co-rotation results in a reduction of the typical speeds of DM particles incident on Earth, and this can have profound effects on the expectations for direct detection experiments \citep{bruch_detecting_2009} \citep[however, see][]{billard_is_2013} and the DM capture rates in the Earth and Sun \citep{sivertsson_accurate_2010}.

\begin{deluxetable*}{lcccccccccc}
\tablewidth{0pt}
\tablecolumns{11}
\tablecaption{Properties of the simulated MW galaxies at $z=0$.\label{tab:MW_prop}}
\tablehead{
MW Halo & $M_{\rm vir}$ & $R_{\rm vir}$ & $V_{\rm peak}$ & $N_{\rm tot}$ & $M_{\rm DM}$ & $M_{\rm gas}$ & $M_\star$ \\
        & [$\msun$]   & [kpc]        & [\kms]        &              & [$\msun$]   & [$\msun$] & [$\msun$]
}
\startdata
ErisDark & $9.1 \times 10^{11}$ & 247 & 166 & $7.55 \times 10^6$ & $9.1 \times 10^{11}$ & 0 & 0 \\
Eris & $7.8 \times 10^{11}$ & 235 & 239 & $1.85 \times 10^{7}$ & $6.9 \times 10^{11}$ & $5.6 \times 10^{10}$ & $3.9 \times 10^{10}$
\enddata
\end{deluxetable*}

In the present paper, we analyze the density and velocity structure of the local DM distribution in one of the highest resolution and most realistic hydrodynamic cosmological calculations of the formation of a disk-dominated galaxy, the Eris simulation. Our work represents an improvement over past analyses based on hydro simulations in at least two aspects. First, Eris represents a simulated disk galaxy that matches, for the first time, many of the observed properties of our Milky Way. Secondly, we compare Eris to its DM-only twin simulation ErisDark, a collisionless run starting from identical initial conditions, which allows us to isolate the effects of the dissipational baryonic physics.
The reader should be cautioned, however, that the total mass of the resulting simulated galaxy ($8 \times 10^{11} \msun$) falls at the lower end of the wide range of estimates for the virial mass of the Galaxy ($5 \times 10^{11} \msun < \mvir < 3 \times 10^{12}$). Moreover, also as a consequence of the low mass, its merger history appears relatively quiescent: within the $\Lambda$CDM cosmology, the fraction of halos of Eris' present-day mass that have not experienced a merger with mass ratio 1:10 or larger since redshift 3 is about 15\% \citep{koda_2009}. While a comparison of the kinematic properties of halo stars in Eris with the latest sample of halo stars from SDSS seems to favor a light, centrally concentrated Milky Way halo \citep{rashkov_light_2013}, no strong, definitive arguments exist as of yet to constrain the timing and mass-ratio abundance of our Galaxy assembly history. Undoubtedly, the presented results will depend quantitatively on the specific halo-assembly realization.

Our paper is organized as follows. In Section~\ref{sec:baryonic_physics}, we review the properties of the simulations Eris and ErisDark, and present the local density and velocity distributions. In Section~\ref{sec:darkdisk} we focus on the dark disk component by analyzing material that was stripped from accreted satellites. Finally, in Section~\ref{sec:implications} we discuss implications for direct detection experiments, including effects on the mean scattering rate and the annual modulation, before summarizing our main results in Section~\ref{sec:summary}.\\

\section{Baryonic Physics and the Local Dark Matter Distribution}\label{sec:baryonic_physics}

In this section we describe the properties of the local (i.e. near 8 kpc in the disk) distribution of dark matter as predicted in Eris, a high resolution cosmological hydrodynamics galaxy formation simulation resulting in a realistic Milky Way analog. In order to elucidate the effects that the dissipational baryonic physics has had, we compare to results from ErisDark, the DM-only counterpart to Eris. We first briefly review the properties of the two simulations, then describe the local DM density and velocity structure.

\subsection{The Eris and ErisDark Simulations}
Both Eris and ErisDark are cosmological zoom-in simulations of a Milky-Way-like galaxy drawn from an N-body realization of 40 millions dark-matter particles in a 90-Mpc-side periodic box. The initial conditions have been generated with the code {\sc grafic1} \citep{bertschinger_multiscale_2001}, assuming first-order Zeld'ovich approximation for the displacements and velocities of the particles at $z_i = 90$, and a $\Lambda$CDM cosmological model ($H_0 = 73 $ km s$^{-1}$ Mpc$^{-1}$, $\Omega_m = 0.268$, $\Omega_b = 0.042$, n$_s = 0.96$, $\sigma_8 = 0.76$). Three levels of refinement have been implemented \citep[e.g.][]{katz_hierarchical_1993} to obtain a high-resolution particle mass of $m^{\rm ErisDark}_{\rm DM} = 1.2\times 10^5 \msun$ in a subregion of about 1 Mpc side. We have modeled the process of structure formation by following two distinct twin runs, Eris and ErisDark, simulated with the N-body+SPH code {\sc gasoline} \citep{wadsley_gasoline:_2004}, respectively with and without including baryonic dynamics and physics. In the Eris SPH simulation, presented in detail in \citet{guedes_forming_2011}, the high-resolution particles are further split into 13 millions dark-matter particles and an equal number of gas particles. The final dark and gas particle mass reads $m_{\rm DM} = 9.8 \times 10^4 \msun$ and $m_{\rm SPH} = 2 \times 10^4 \msun$, while each star particle is stochastically created with an initial mass of $m_\star = 6.1 \times 10^3 \msun$. The gravitational softening length is fixed to 124 physical pc at all redshifts $z < 9$, both in ErisDark and Eris. Compton, atomic, and metallicity-dependent radiative cooling at low temperatures, heating from a cosmic UV field and supernova explosions, a star formation recipe based on a high atomic gas density threshold (n$_{\rm SF} = 5$ atoms cm$^{-3}$ with 10 percent star-formation efficiency), and a blastwave scheme for supernova feedback give rise at the present epoch to a massive, barred, late-type spiral galaxy, a close Milky Way analog \citep{rashkov_light_2013}.

The basic properties of the two simulated Milky-Way galaxies at $z=0$ are summarized in Table \ref{tab:MW_prop}, where the host halo has been identified with the spherical overdensity Amiga Halo Finder \citep{gill_evolution_2004,knollmann_ahf:_2009} (the virial radius being defined to enclose a mean total density of 98 times the critical density of the Universe at a given time, i.e. 364 $\bar\rho$, Bryan et al. 1998\nocite{bryan_statistical_1998}). In Eris, the simulated galaxy has an extended rotationally supported stellar disk with a radial scale length R$_d$ = 2.5 kpc, a scale height of $h_{z}$ = 490 pc at a galactocentric distance of 8 kpc, a gently falling rotation curve, with a circular velocity at the solar circle of $V_{c,\odot}=205 \kms$, in good agreement with the recent determination of the local circular velocity, $V_{c,\odot}=218 \pm 6 \kms$ by \citet{bovy_milky_2012}, an i-band bulge-to-disk ratio B/D = 0.35, and a baryonic mass fraction within the virial radius that is 30 percent below the cosmic value. The stellar mass contained in the thin disk is $\sim 2 \times 10^{10} \msun$, 50 percent of the overall stellar content at $z=0$. An in-depth description and discussion of the structural properties, brightness profiles, stellar and gas content in Eris can be found in \citet{guedes_forming_2011}.

\subsection{Density profiles}\label{sec:density}

\begin{figure}[htp]
\includegraphics[width=\columnwidth]{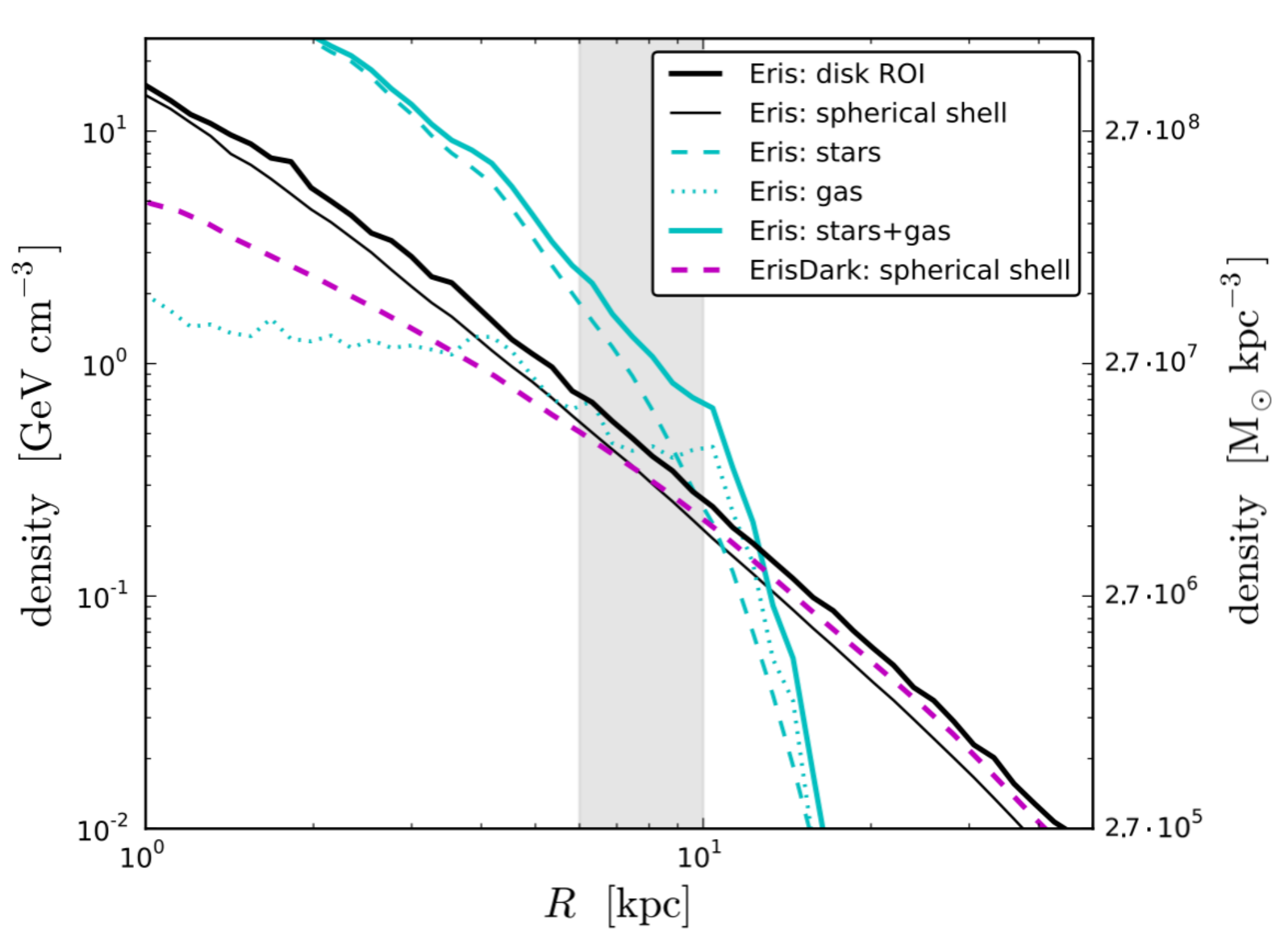}
\caption{Density profiles as a function of galactocentric radius in the midplane ($|z| <$ 0.1 kpc) of the stellar disk. The thick black line is for all DM particles in the disk region,
and the cyan lines are the baryonic components. The thin black line is the spherically-averaged dark matter density profile, for which $R$ refers to the 3D radius, and the magenta dashed line is the same for the ErisDark simulation. The shaded band indicates the region we considered for the velocity distribution analysis.}
\label{fig:density_profile}
\end{figure}

In Figure~\ref{fig:density_profile} we show the density profiles of all matter components (dark matter, stars, and gas) as a function of galactocentric radius. We define a disk region-of-interest (ROI) as a cylindrical volume aligned with the stellar disk and extending 0.1 kpc above and below the disk's midplane ($|z|<0.1$ kpc), and calculate density profiles by binning up particles in evenly spaced logarithmic cylindrical annuli. We show profiles for all DM particles contained in this ROI (thick black line), as well as the baryonic components (cyan lines). For comparison, we also plot a spherical density profile obtained by binning in spherical shells all DM particles in Eris (thin black line) and ErisDark (dashed magenta line).


The Eris galaxy is baryon dominated inward of 12.5 kpc. DM makes up only slightly more than half (55.5\%) of the enclosed mass within a spherical radius of 8 kpc, implying that the circular velocity at 8 kpc is sourced in about equal parts by DM and baryons. The local DM density at 8 kpc in the disk plane is 0.42 GeV cm$^{-3}$ (spanning between 0.82 and 0.27 GeV cm$^{-3}$ in the 6-10 kpc range) and it contributes only 27.5\% of the total matter density at this radius. The most recent observational constraints on the local DM density span from $1.25^{+0.30}_{-0.34}$ GeV cm$^{-3}$ \citep{garbari_limits_2011} to $0.3 \pm 0.1$ GeV cm$^{-3}$ \citep{bovy_local_2012}, with large uncertainties due to modeling assumptions: in this respect, Eris' local DM density appears in good agreement with observationally inferred estimates. The total baryonic content in Eris' ROI (spanning between 2.7 and 0.6 GeV cm$^{-3}$ in the 6-10 kpc range) appears lower than the results from the \textit{Hipparcos} satellite reported by \cite{holmberg_localdensity_2000}, 
who derive an estimate of the local dynamical mass density of
$0.1 \msun$ pc$^{-3} = 3.75$ GeV cm$^{-3}$ to be compared with the measurement of $0.095 \msun$ pc$^{-3}$ = 3.56 GeV cm$^{-3}$ in 
visible disk matter only \footnote{The total baryonic density in Eris' disk declines by about 40\% by varying the ROI height from 1 to 4 times the force resolution, where 0.490 kpc is our estimate for the scale height of Eris' disk. On the other hand, the local DM density is insensitive to the choice of the ROI height, up to $|z| <$2 kpc: this suggests already that if a dark disk can effectively be identified, its vertical extension will be much larger than the baryonic disk's.}.
While this tension depends on the effective thickness of Eris' baryonic disk (still inevitably puffed up compared to the Milky Way because of resolution), it should be noticed that the total surface density for $|z| <$1.1 kpc at 8 kpc (48 $\msun$pc$^{-2}$) is remarkably consistent with the range of local surface densities recently derived by \cite{bovy_rix_2013}.

Interestingly, the local DM density in the disk is about 31\% \textit{higher} than the spherically averaged DM density at 8 kpc in the ErisDark simulation (0.32 GeV cm$^{-3}$), even though in ErisDark all of the matter is treated as DM, while in Eris 17\% is baryonic. This increase in the local DM density is the result of a contraction due to the dissipational processes occurring during the formation of the Galactic disk. The local disk DM density in Eris is also higher (by 34\%) than its spherical average (0.31GeV cm$^{-3}$), indicating that at 8 kpc this contraction occurred primarily in the plane of the disk rather than globally. 

This conclusion is further strengthened by a comparison of the ellipsoidal shapes of the dark matter density distributions in Eris and ErisDark. We followed the iterative method described in \citet{kuhlen_shapes_2007} and applied it to particles between 6 and 10 kpc from the host halo's center. As is typical for halos in dissipationless DM-only simulations \citep[e.g.][]{allgood_shape_2006}, the ErisDark halo is quite prolate, with intermediate-to-minor axis ratio $q=0.53$ and minor-to-major axis ratio $s=0.45$. As expected \citep{katz_dissipational_1991,dubinski_effect_1994,kazantzidis_effect_2004,abadi_galaxy-induced_2010}, the inclusion of dissipational baryonic physics results in a more axisymmetric and rounder DM halo in Eris. It is oblate with $q=0.99$ and $s=0.69$, and its minor axis is aligned to within $1.5^\circ$ with the angular momentum vector of the stellar disk (and to within $7^\circ$ of ErisDark's minor axis).

\subsection{Velocity Distributions}

\begin{figure*}[htp]
\begin{center}
\includegraphics[width=17cm]{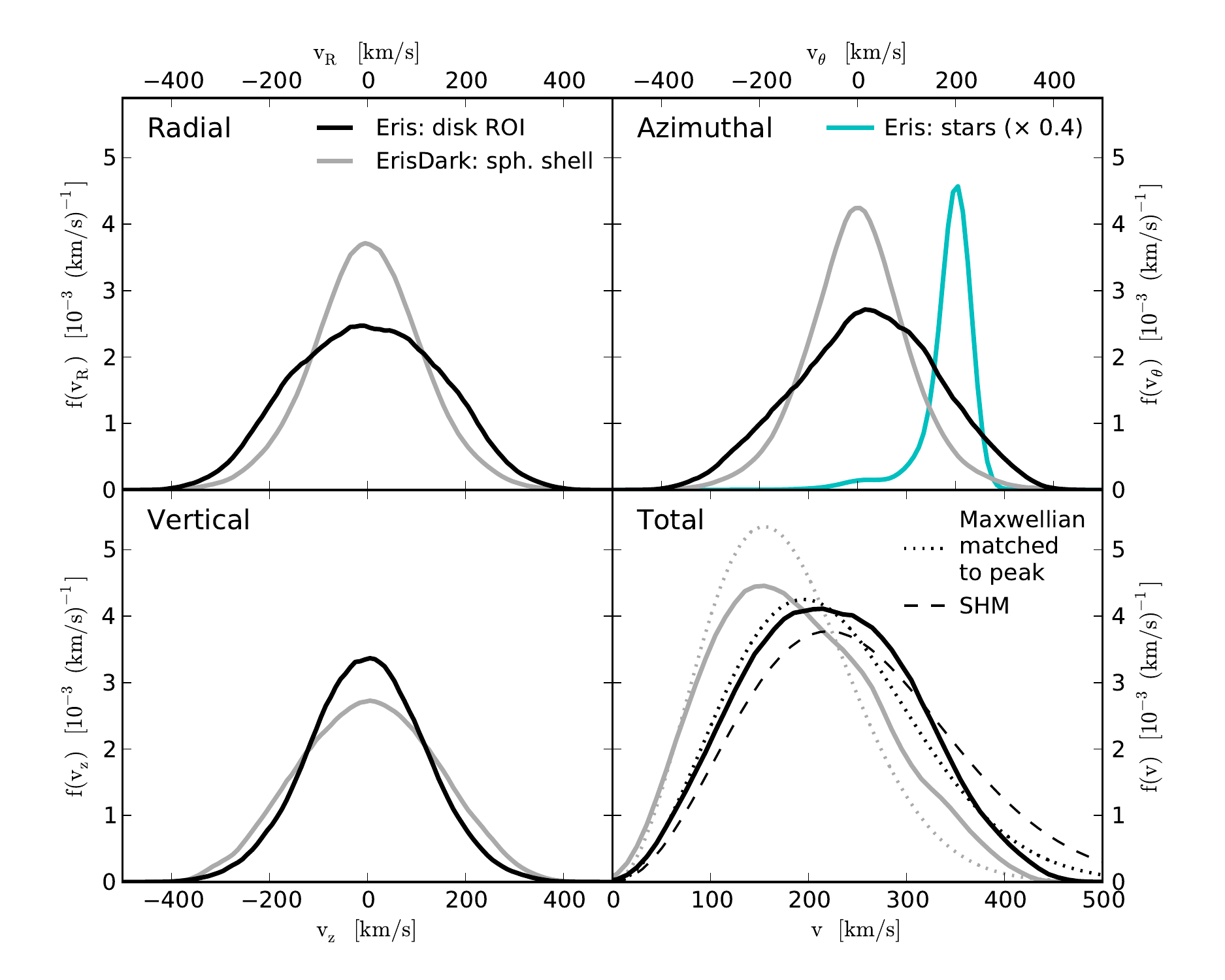}
\caption{DM velocity distributions in the Galactic rest frame for particles in an annulus near the Sun's location ($R_\odot =$ 8 kpc): radial (top left), azimuthal (top right), vertical (bottom left) components, and the velocity modulus (bottom right). For Eris (black) we show distributions for particles in the disk ($|R - R_\odot| <$ 2 kpc, $|z| <$ 2 kpc), while for ErisDark (grey) all particles within a spherical shell of width 4 kpc are used. In the upper right we additionally show the distribution of Eris star particles (cyan, scaled by a factor of 0.4). In the lower right, we also show Maxwellian curves (dotted) with the same peak speed as the simulations' distribution ($\vpeak = 195 \kms$ in Eris and $155 \kms$ in ErisDark), as well as the Standard Halo Model with $\vpeak = 220 \kms$ (dashed). The simulation curves have been smoothed with a boxcar window of width $50 \kms$. }
\label{fig:fv_4panel}
\end{center}
\end{figure*}

Scattering rates at DM direct detection experiments depend on the shape of the DM velocity distribution $f(\vec{v})$ at Earth. The relevant length scale ($R_\oplus$) is far below the resolution limit even of state-of-the-art numerical galaxy formation simulations (few 100 parsec), so we are forced to take a coarse-grained spatial average. In ultra-high resolution purely collisionless (DM only) simulations, spatial variations in $f(\vec{v})$ at 8 kpc have been investigated on $\sim$ kpc scales \citep{vogelsberger_phase-space_2009,kuhlen_dark_2010}. These studies found some spatially localized sharp velocity features due to the presence of subhalos or tidal streams, but only with a low probability of $\sim 10^{-2}$. For the present work we thus neglect any small scale variations and consider the velocity distribution determined from all particles in a cylindrical disk annulus to be representative of $f(\vec{v})$ at Earth.

The annulus we consider is aligned with the stellar disk and has $|R - R_\odot| <$ 2 kpc and $|z| <$ 2 kpc.\footnote{Note that we have considerably extended the vertical extent of the annulus ROI compared to the disk ROI used for the density profiles (Sec.~\ref{sec:density}). This is necessary in order to get particle numbers sufficient to determine velocity distribution.} In Eris this region contains 81,213 DM particles and 830,068 star particles. From these we calculate distributions of the radial (\vR), azimuthal (\vtheta), and vertical (\vz) velocity components, as well as for the velocity modulus ($|\vec{v}|$). These distributions are shown in Figure~\ref{fig:fv_4panel}. All distributions are separately normalized to unity ($\int \! f(v_i) \, dv_i = 1$), and have been smoothed with a boxcar window of width $50 \kms$ in order to suppress numerical noise stemming from low particle counts. The distribution of the star's $\vtheta$ (cyan line in upper left panel) has been scaled down by a factor of 0.4 in order to show its shape on the same plot.

We compare the Eris disk ROI velocity distributions to the ErisDark spherical shell sample of width 4 kpc, which contains 229,931 DM particles. This kind of spherical shell sample is commonly used in the analysis of DM-only simulations of Milky-Way-like halos, for which there is no preferred plane to associate with the Galactic disk. We additionally plot a Maxwell-Boltzmann (MB) distribution with the same peak speeds as the simulations' distributions:  $\sigma_{\rm 1D} = v_{\rm peak}/\sqrt{2} = 137.9 \, (109.6) \kms$ in Eris (ErisDark).

Compared to ErisDark, the dissipational baryonic physics in Eris has broadened the radial and azimuthal distributions, while the vertical component has become slightly narrower. Note that the azimuthal component in Eris is skewed towards positive $\vtheta$, indicating the presence of an enhanced population of particles approximately co-rotating with the stars, i.e. a so-called ``dark disk''. This asymmetry is the topic of Section~\ref{sec:darkdisk}. 

In the speed distribution (lower right), the DM-only simulation exhibits the familiar departures from a Maxwellian shape \citep{hansen_universal_2006,vogelsberger_phase-space_2009,kuhlen_dark_2010}, with a deficit near the peak and excess particles at high speeds.  In Eris the distribution is shifted to larger speeds, with the mean speed increasing from $\langle v \rangle = 187.6 \kms$ to $220.8 \kms$. Furthermore, it no longer shows as marked a departure from the matched Maxwellian as in the DM-only case, only exceeding it slightly from 230 to 380 \kms and falling more rapidly at even higher speeds. We also compared to the so-called Standard Halo Model (SHM) distribution, consisting of a Maxwellian with $\vpeak = 220 \kms$ (dashed line). Eris actually exceeds the SHM at all speeds less than $\sim 350 \kms$, and then again falls more sharply at higher speeds. 

\begin{figure}[htp]
\includegraphics[width=\columnwidth]{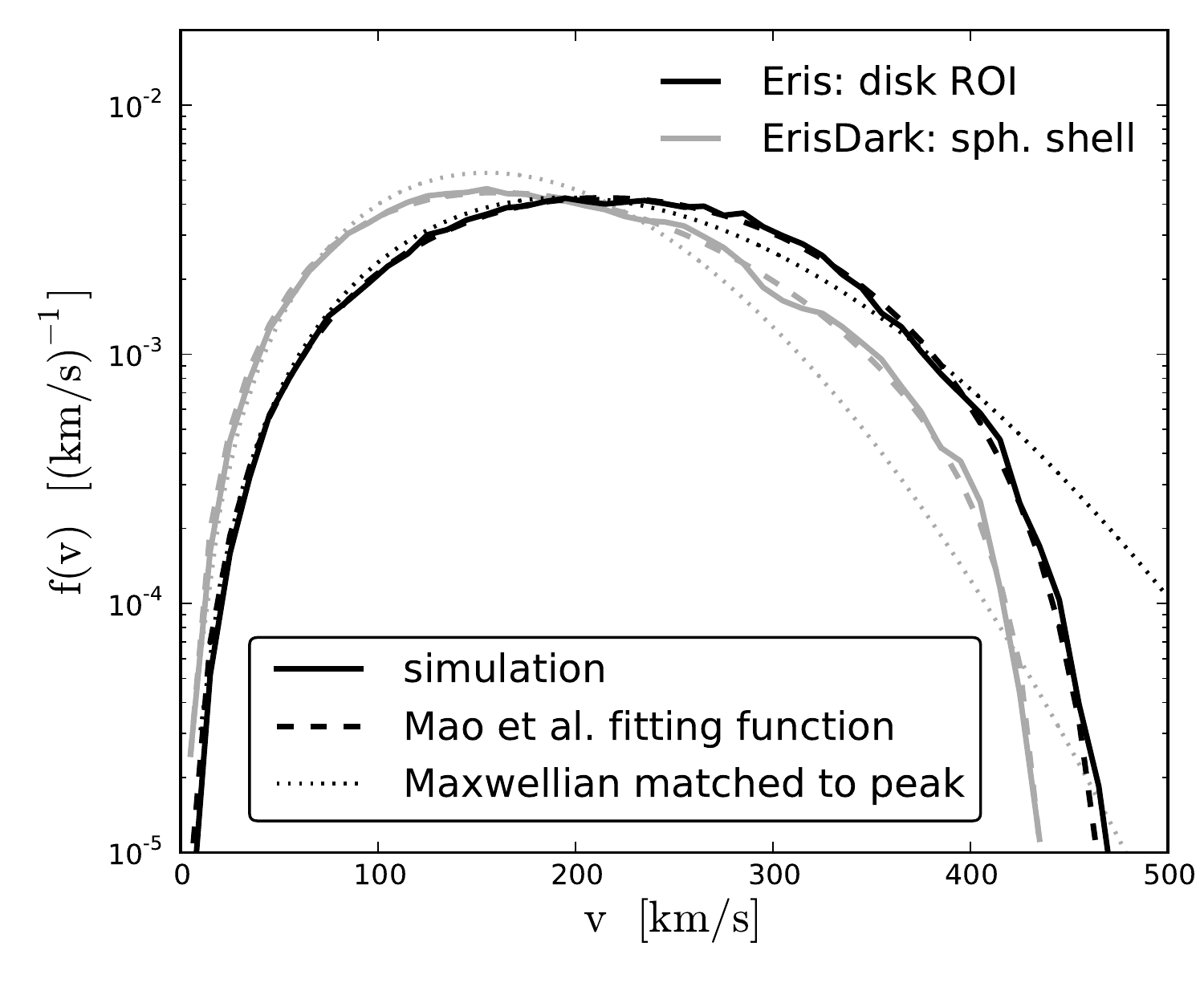}
\caption{The speed distribution in Eris (black) and ErisDark (grey) on a logarithmic scale, compared to the fitting function from \citet{mao_halo--halo_2013} (dashed), with $(v_0,v_{\rm esc},p) = (330 \kms, 480 \kms, 2.7)$ for Eris and $(100 \kms, 440 \kms, 1.5)$ for ErisDark, and to the peak-matched Maxwellian curves (dotted).}
\label{fig:Mao_comparison}
\end{figure}

Recently \citet{mao_halo--halo_2013} proposed an empirical fitting function for the speed distribution\footnote{Note that we include the factor of $4\pi v^2$ in our definition of \fv\ (such that $\int \! f(v) dv = 1$), and hence our expression has an additional factor of $v^2$ compared to \citet{mao_halo--halo_2013}.},
\begin{equation}
f(v) =
\begin{cases}
A \, v^2 \, \exp\!{(-v/v_0)} \, \left( \vesc^2 - v^2 \right)^p & \text{if $v \leq \vesc$,} \\
0 & \text{otherwise,}
\end{cases}
\end{equation}
which they showed to be flexible enough to match the variations in the shape of \fv\ over a wide range of halo masses and locations within the halos. As shown in Figure~\ref{fig:Mao_comparison}, the \fv\ of both ErisDark and Eris are indeed well fit by this functional form, with fit parameters $(v_0, p) = (330 \kms, 2.7)$ in Eris and $(100 \kms, 1.5)$ in ErisDark. The escape velocity \vesc\ is not a free parameter and was determined directly in the simulations from the maximum particle speeds in the ROI to be $\vesc = 480 \kms$ in Eris and $440 \kms$ in ErisDark.

The increase in the parameter $p$ from ErisDark to Eris is an important result of this study, since it is precisely such high values, i.e. a more steeply falling \fv\ at high speeds, that ease the tension between the tentative detection of a scattering signal reported by CDMS-Si \citep{cdms_collaboration_dark_2013} and the nominal exclusion of such a signal from the Xenon-100 experiment \citep{aprile_dark_2012}, as shown by \citet{mao_connecting_2013}.\\

\begin{figure*}[htp]
\includegraphics[width=\textwidth]{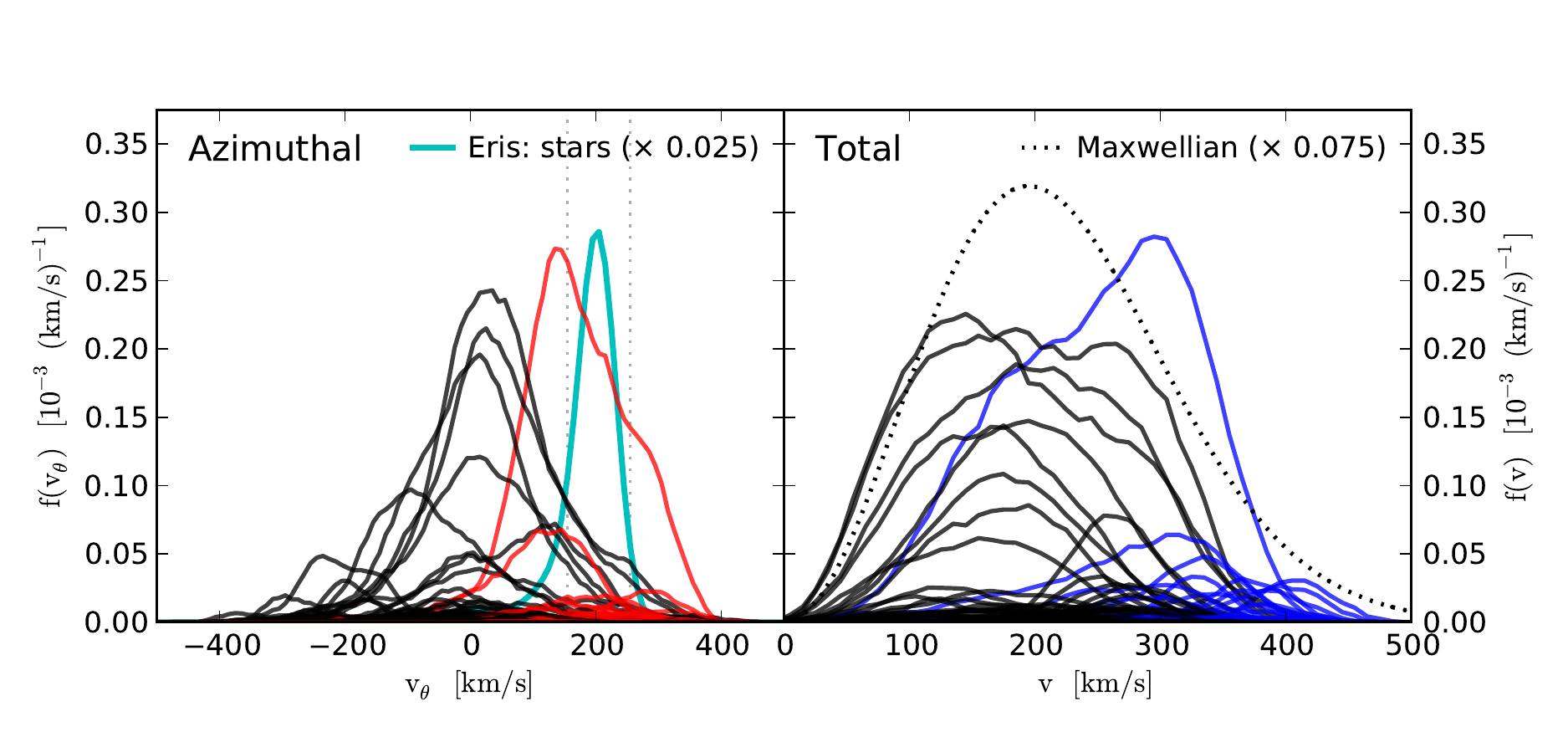}
\caption{The contributions to the azimuthal (left) and total (right) velocity distributions from accreted satellites. In the left panel, the red lines are satellites with high asymmetry parameter (${\rm Asym} > 2/3$, see Eq.~\ref{eq:Asym} and main text for details) and the cyan line is the stellar distribution (scaled by a factor of 0.025). The two vertical dotted lines mark the range of \vtheta\ commonly used in the definition of a ``dark disk'' sample: $|\vtheta - v_{\rm peak, stars}| < 50 \kms$. In the right panel, the blue lines denote satellites contributing to a high speed ``debris flow'' ($\vpeak > 250 \kms$) and the dotted line is a Maxwellian matched to the peak of the full DM distribution ($\vpeak = 195 \kms$, scaled by a factor of 0.075).}
\label{fig:fv_2panel_darkdisk}
\end{figure*}

\section{A rotating dark disk from satellite accretion}\label{sec:darkdisk}

We now turn to a discussion of the origin of the asymmetry in the distribution of the azimuthal velocity component in Eris. This feature is indicative of a ``dark disk'', consisting of an oblate dark matter distribution aligned and nearly co-rotating with the stellar disk, as previously reported in cosmological hydrodynamic galaxy formation simulations by, for example, \citet{read_dark_2009} \citep[see also][]{ling_dark_2010}, who find that the dark disk may contribute between 20 and 60 percent of the local DM density. 

Dark disk material is typically thought to be deposited by massive satellites that are preferentially dragged into and disrupted in the plane of the disk \citep{read_thin_2008,read_dark_2009}. If such satellites more commonly have pro-grade orbits with respect to the rotation of the stars in the disk, then the material they deposit upon being tidally disrupted would predominantly be co-rotating with the stars, leading to a positively skewed $f(\vtheta)$.
In fact, dark disk material is expected to be prograde rather the retrograde, as dynamical friction is more efficient at dragging towards the disk plane incoming satellites which not only are massive, but also on a prograde orbit wrt to the baryonic disk.

Indeed, the fraction of DM with \vtheta\ within 50 \kms\ of the peak of the star's $f(\vtheta)$ (at 205 \kms) is higher in Eris (0.13) than in ErisDark (0.055). However, a similar increase is also observed at negative \vtheta\ (within 50 \kms\ of $\vtheta = -205 \kms$) where the fraction of mass increases from 0.054 (ErisDark) to 0.095 (Eris). This indicates that the bulk of the increase in co-rotating material stems from a broadening of the $f(\vtheta)$ distribution, rather than from disrupted satellites preferentially depositing material into a co-rotating configuration. Note that some symmetric broadening could still arise from satellites being dragged into the galactic plane, if they were about equally likely to be on prograde and retrograde orbits. Most of the increase in dispersion, however, is likely due to the additional baryonic material that has been able to settle to the center of the halo and has deepened the potential: at 8 kpc, the circular velocity has increased from 140 \kms\ in ErisDark to 205 \kms\ in Eris.

Motivated by these considerations, we have followed in Eris the accretion and disruption of all satellites consisting of more than 1000 DM particles at infall. There are 160 such systems, of which 74 deposit material in our disk ROI. We have identified all DM particles that were at one point bound to these satellites, and determined their contribution to the DM distribution in the disk ROI. In total there are 30,780 such particles, together contributing 38 percent of the local DM density. Figure~\ref{fig:fv_2panel_darkdisk} shows distributions of the azimuthal and total speed for each of these satellites. The left panel demonstrates that not all of the material accreted from satellites is co-rotating with the stars. In fact, most satellites deposit material into a roughly symmetric \vtheta\ distribution centered on $\vtheta = 0 \kms$, while a few deposit a predominantly retrograde ($\vtheta < 0 \kms$) particle distribution. The positive skewness in the total $f(\vtheta)$ appears to be contributed mostly by one massive system with $M_{\rm infall} = 1.8 \times 10^{10} \msun$ and $z_{\rm infall} = 2.7$ (mass ratio 1:14). 

The total speed distributions in the right panel reveals two populations of accreted satellites: one set deposits material with typical speeds comparable to the peak of the overall speed distribution ($\sim 200 \kms$), and a second set with considerably higher speeds ($\gtrsim 300 \kms$). The latter material makes up a so-called ``debris flow'', whose implication for direct detection experiments has been discussed in \citet{kuhlen_direct_2012}.

\begin{figure}[htp]
\centering
\includegraphics[width=\columnwidth]{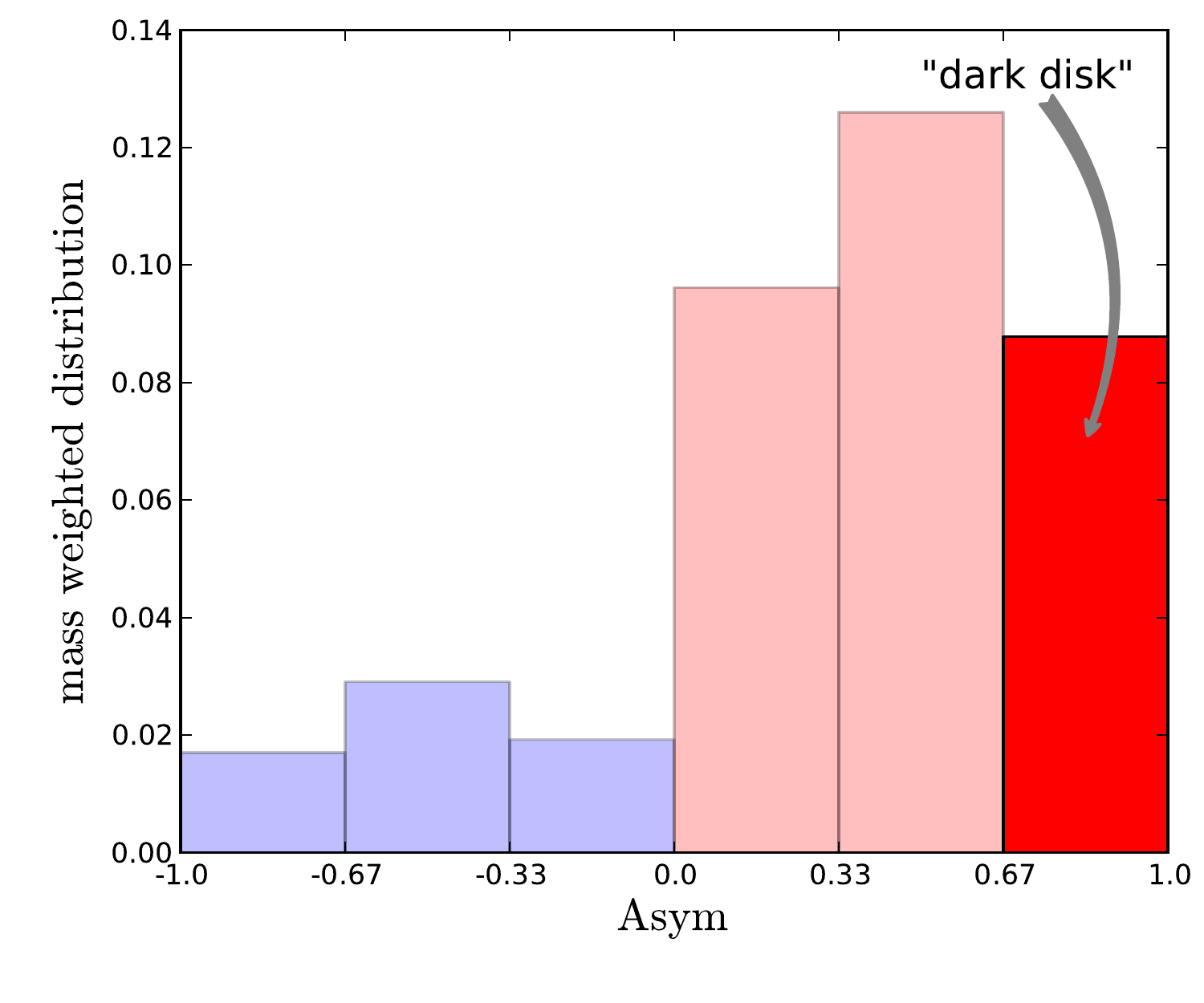}
\caption{Histogram of the asymmetry parameter Asym (see text for definition) of accreted satellites, weighted by the mass they contribute locally. We define as a ``dark disk'' material contributed by satellites with ${\rm Asym} > 2/3$.}
\label{fig:Asym_histogram}
\end{figure}

Returning to the azimuthal distributions, we define for each satellite an asymmetry parameter,
\begin{equation}
{\rm Asym} = \f{F(\vtheta > 0) - F(\vtheta < 0)}{F(\vtheta > 0) + F(\vtheta < 0)},
\label{eq:Asym}
\end{equation}
where $F(\vtheta > 0) = \int_0^\infty \! f(\vtheta) d\vtheta$ is the fraction of material with positive \vtheta, and $F(\vtheta < 0)$ the fraction with negative \vtheta.
This parameter quantifies the degree to which a satellite contributes material predominantly rotating in the same sense as the stars. Figure~\ref{fig:Asym_histogram} shows a histogram of the parameter Asym, weighted by the mass contributed to the disk ROI by each satellite. The majority of the mass accreted from satellites has positive Asym, rotating prograde with respect to the stars. $\Asym > 0$ satellites contribute 31 percent of the DM mass in the disk ROI, 81\% of all accreted material. 

Following previous work \citep{purcell_dark_2009,read_dark_2009}, we first consider as dark disk particles those with \vtheta\ within 50 \kms\ of the peak of the stellar $f(\vtheta)$ at 205 \kms. However, even satellites with a fairly symmetric $f(\vtheta)$ can contribute material that satisfies this criterion; indeed, an almost equal amount of mass is found to be rotating in the opposite sense at the same speed. For these reasons, we additionally impose the constraint that $\Asym > 2/3$, i.e. only material from satellites with highly positively asymmetric $f(\vtheta)$ is considered part of the dark disk. With these criteria, the dark disk in Eris makes up only 3.2 percent ($2.6 \times 10^8 \msun$) of all DM in the disk ROI. Note that this component almost exactly accounts for the difference in mass between material with \vtheta\ within 50 \kms\ of 205 \kms\ and -205 \kms\ ($2.75 \times 10^8 \msun$).


With the above definition, the dark disk in Eris is much less massive than what has been suggested by previous work \citep{read_thin_2008, read_dark_2009, purcell_dark_2009, ling_dark_2010}. A physical reason for this difference is the fairly quiet satellite accretion history of Eris, whereas previous works have focused on Milky-Way galaxies characterized by more massive and more numerous merger events. It has previously been pointed out that a quiet recent merger history is likely a pre-requisite for obtaining a realistic Milky Way analog with a thin and cold stellar disk, thus making a heavy dark disk unlikely \citep{purcell_dark_2009}. However, other studies have shown that the over-heating of the thin stellar disk is much less disruptive once more realistic distributions for the inclination and eccentricities of satellite orbits \citep{read_thin_2008} and gas \citep{moster_gasdisk_2010} are properly included in the simulations. Moreover, \cite{niederste-ostholt_sagittarius_2010} suggest that Sagittarius, which is currently being disrupted in the Galaxy' s tidal field, might have been as massive as $10^{10} \msun$ prior to merging, making the necessity of a quiescent merger history even more uncertain. Our $\Asym > 2/3$ cut further reduces the dark disk contribution.

Although the dark disk, under the above criteria, only contributes 0.032 of the total DM mass in the disk ROI, we note that the overall asymmetry in $f(\vtheta)$ is considerably larger: $\Asym = 0.12$, implying about 30\% more prograde than retrograde material. This motivates a less restrictive definition of dark disk, namely all material contributed by satellites with high asymmetry, regardless of its lag speed with respect to the stars. When only applying the asymmetry criterion of $\Asym > 2/3$, the dark disk makes up 9.1 percent ($7.25 \times 10^8 \msun$) of all DM in the disk ROI, and it is this definition of dark disk that we use in the remainder of the paper. About 60\% of the dark disk is contributed by the same single satellite that dominates the overall skewness of $\fv$. 

\begin{figure}[htp]
\centering
\includegraphics[width=0.95\columnwidth]{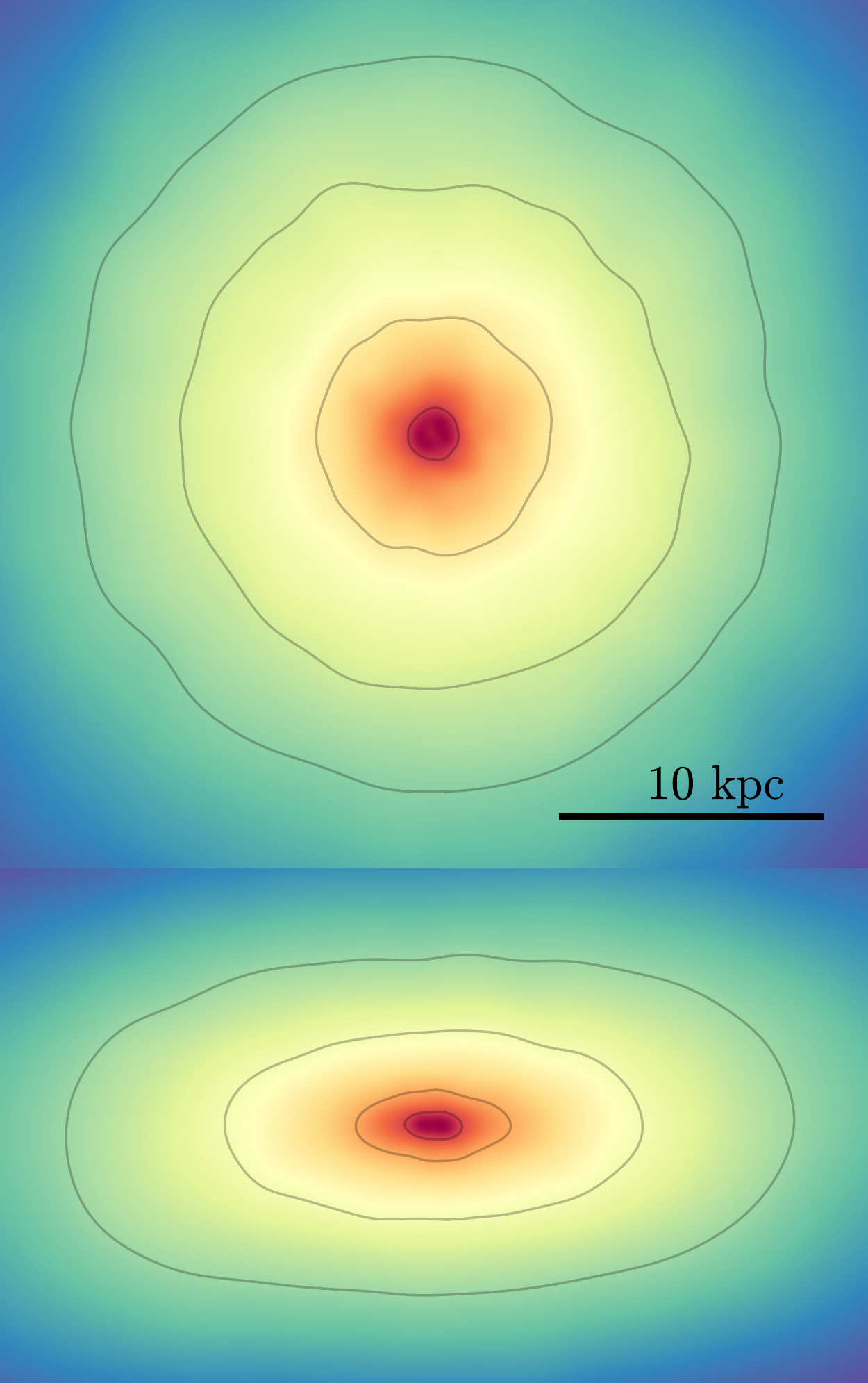}
\caption{Face-on and edge-on projections of the Eris dark disk (Asym $> 2/3$) particles. Contours are at $(6.7 \times 10^6, \, 1.0 \times 10^7, \, 2.0 \times 10^7, \, 6.7 \times 10^7) \, \msun \, {\rm kpc}^{-2}$ in the top panel and at $(1.3 \times 10^7, \, 2.6 \times 10^7, \, 6.7 \times 10^7, \, 1.3 \times 10^8) \, \msun \, {\rm kpc}^{-2}$ in the bottom.}
\label{fig:darkdisk_projections}
\end{figure}

\begin{figure}[htp]
\includegraphics[width=\columnwidth]{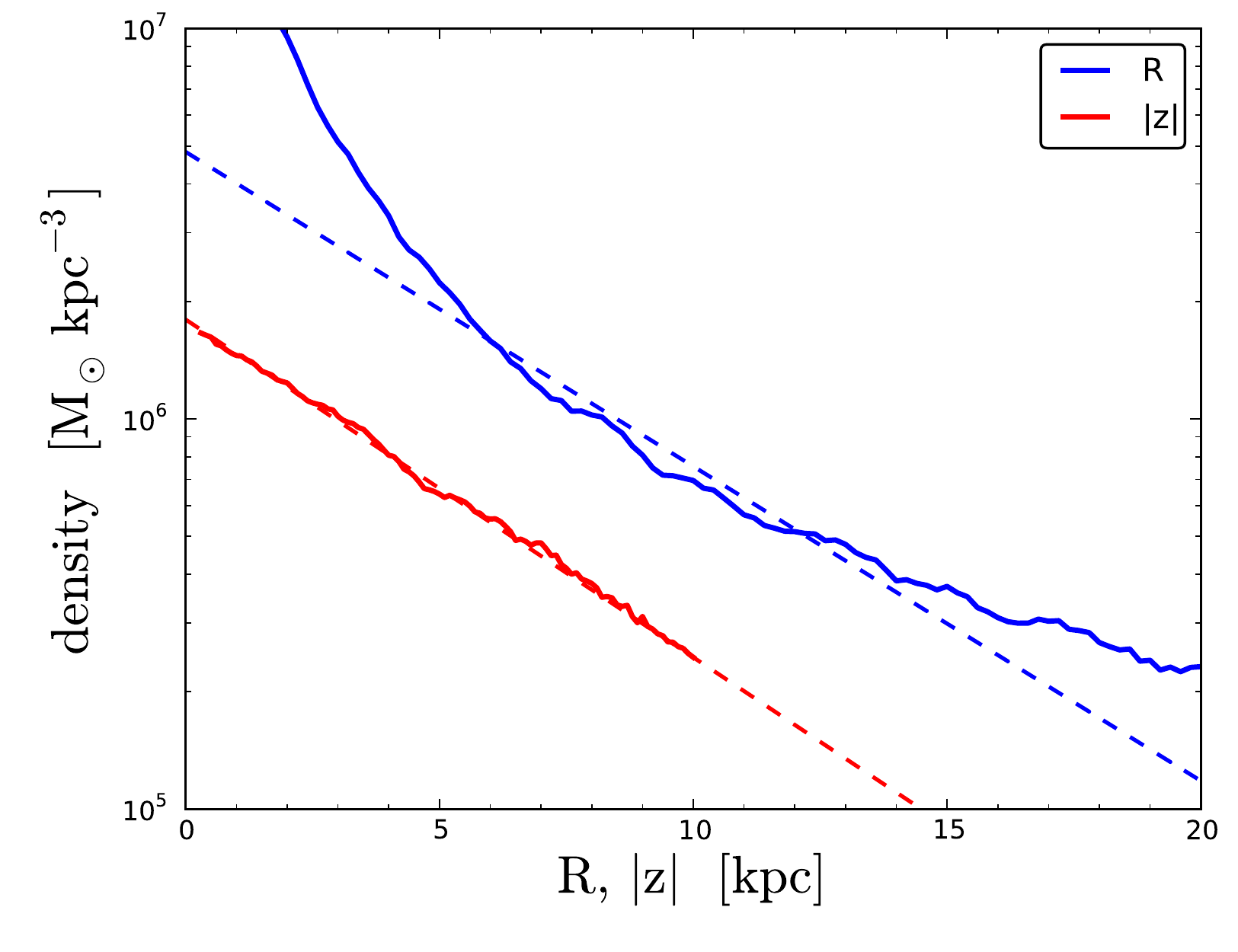}
\caption{Radial (blue) and vertical (red) density profiles of the dark disk component in Eris. The dashed line are exponential profiles with scale radius $R_{\rm dd} = 5.0$ kpc and scale height $h_{\rm z,dd} = 5.4$ kpc. }
\label{fig:exponential_disk}
\end{figure}

Figure~\ref{fig:darkdisk_projections} depicts face-on and edge-on projections of the dark disk material. This exhibits very little azimuthal structure and, as expected, is oblate in shape, with a minor-to-major axis ratio of $s = 0.45$, so even more flattened than the overall DM distribution. As shown in Figure~\ref{fig:exponential_disk}, the vertical density structure of the dark disk is well described by an exponential profile with a scale height of 5.0 kpc. While the radial structure is not exponential over the full radial range, it is approximately so near the solar radius (6--12 kpc), with a scale radius of 5.4 kpc.

The dark disk contributes 0.034 GeV cm$^{-3}$ to the DM density in the disk ROI. This is only about one third of the excess DM density in the Eris disk ROI over the ErisDark spherical average (0.12 GeV cm$^{-3}$), and this suggests that there are at least two distinct processes leading to an enhancement of the DM in the disk plane: one process that results in a DM component with significant net angular momentum and that is nearly co-rotating with the stellar disk, and another process that pulls DM into the disk plane without forcing it to co-rotate \citep[see also][]{zemp_impact_2012}.

In the following section we look in more detail at how the baryonic physics effects we have discussed above affect the expected direct detection signals.\\

\section{Implication for Experiments}\label{sec:implications}

\subsection{Earth Frame $f(\vtheta)$ and \fv}

\begin{figure*}[htp]
\includegraphics[width=\textwidth]{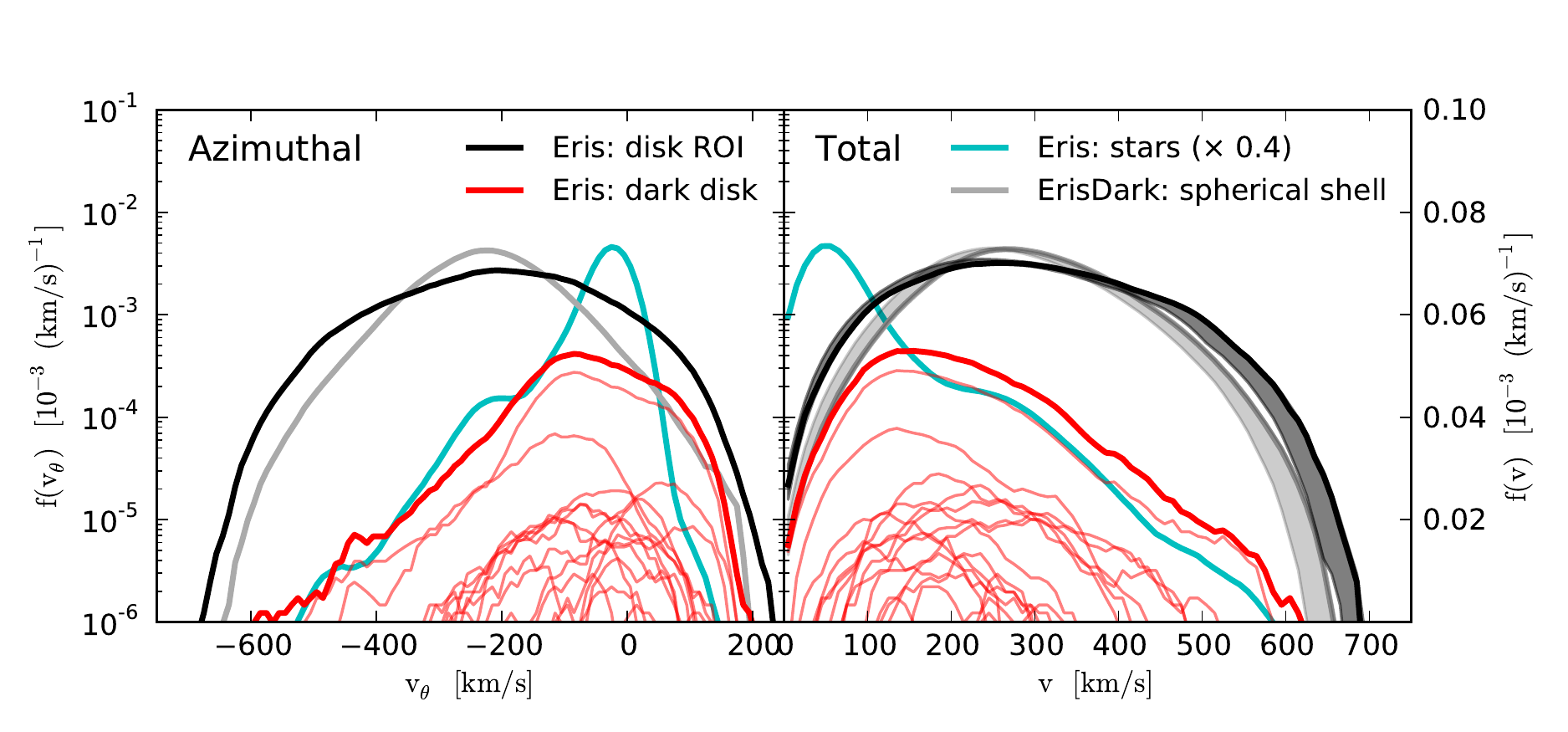}
\caption{The azimuthal (left) and total (right) velocity distributions in the Earth rest frame (June 1). The line styles are as in Figure~\ref{fig:fv_4panel}, and additionally we show the total ``dark disk'' contribution (thick red) and that of individual Asym $> 2/3$ satellites (thin red). The gray shaded regions indicate the extent of the annual modulation in the total $\rm f(v)$.}
\label{fig:fv_earth_2panel_darkdisk}
\end{figure*}

\begin{deluxetable*}{lccccc}
\tablewidth{0pt}
\tablecolumns{6}
\tablecaption{Table of experimental benchmarks.\label{tab:benchmarks}}
\tabletypesize{\normalsize}
\tablehead{
Experiment & Target (Z, A) & $m_\chi$ & $E_r$ & \vmin & Reference \\
           &               &   [GeV] & [keV$_{\!\rm nr}$] &     [km s$^{-1}$] &
}
\startdata
   CDMS II (Ge) &    Ge (32, 73) &      10 &   $[10.0, 100]$ & $[651, 2060]$ & (1) \\[2pt]
          &               &      70 &                & $[160, 507]$ &  \\[2pt]
          &               &     500 &                & $[89.9, 284]$ &  \\[2pt]
\tableline\\[-5pt]
   CDMS II (Si) &    Si (14, 28) &      5 &   $[7.0, 100]$ & $[700, 2640]$ & (2) \\[2pt]
          &               &     10 &                & $[403, 1520]$ &  \\[2pt]
          &               &     20 &                & $[254, 961]$ &  \\[2pt]
\tableline\\[-5pt]
  XENON100 &   Xe (54, 131) &      10 &  $[6.6, 43.3]$ & $[671, 1720]$ & (3) \\[2pt]
          &               &     50 &                & $[172, 442]$ &  \\[2pt]
          &               &     500 &                & $[60.0, 154]$ &  \\[2pt]
\tableline\\[-5pt]
DAMA/LIBRA &    Na (11, 23) &      10 &  $[6.7, 20.0]$ & $[378, 652]$ & (4) \\[2pt]
          &    I (53, 127) &     100 &  $[25.0, 75.0]$ & $[214, 370]$ &  \\[2pt]
\tableline\\[-5pt]
    CoGeNT &    Ge (32, 73) &       5 &  $[2.27, 11.2]$ & $[583, 1300]$ & (5) \\[2pt]
          &               &      10 &                & $[310, 689]$ &  \\[2pt]
\tableline\\[-5pt]
    CRESST-II &      O (8, 16) &      10 &  $[12.0, 40.0]$ & $[477, 872]$ & (6) \\[2pt]
          &    Ca (20, 40) &      10 &                & $[581, 1060]$ &  \\[2pt]
          &    W (74, 184) &      50 &                & $[253, 463]$ & 
\enddata
\tablecomments{We give the range of \textit{nuclear} recoil energy ($E_r$ in keV$_{\!\rm nr}$) that each experiment is sensitive to. For experiments that publish recoil energy sensitivity ranges in \textit{electron equivalent} energy ($E_{\rm ee}$ in keV$_{\rm ee}$), we convert these to $E_R$ via $E_{ee} = q \, E_r^{\;x}$, with $(q,x) = (0.3,1)$ for Na (DAMA/LIBRA), $(0.08,1)$ for I (DAMA/LIBRA), and $(0.2,1.12)$ for Ge (CoGeNT).}
\tablerefs{(1) \citet{cdms_ii_collaboration_dark_2010}, (2) \citet{cdms_collaboration_dark_2013}, (3) \citet{aprile_first_2010}, (4) \citet{bernabei_new_2010}, (5) \citet{aalseth_cogent:_2012}, (6) \citet{angloher_results_2012}}
\end{deluxetable*}

So far we have focused on velocity distributions in the halo rest frame, but direct detection scattering rates of course depend on the velocity distribution in the Earth's rest frame. Here the low relative velocity with respect to the stars of the rotating dark disk component can lead to pronounced changes compared to the non-rotating DM. We transform the halo-centric velocities into Earth rest frame by applying a Galilean boost by  $\vec{v}_\oplus(t)$. The Earth's velocity with respect to the Galactic center is the sum of the local standard of rest (LSR) circular velocity around the Galactic center, the Sun's peculiar motion with respect to the LSR, and the Earth's orbital velocity with respect to the Sun,
\begin{equation}
\vec{v}_\oplus(t) = \vec{v}_{\rm LSR} + \vec{v}_{\rm pec} + \vec{v}_{\rm orbit}(t).
\end{equation}
We set $\vec{v}_{\rm LSR} = (0, 205, 0) \kms$ , $\vec{v}_{\rm pec} = (10.00, 5.23, 7.17) \kms$ \citep{dehnen_local_1998}, and $\vec{v}_{\rm orbit}(t)$ as specified in \citet{lewin_review_1996}. The velocities are given in the conventional $(U,V,W)$ coordinate system where $U$ refers to motion radially inwards towards the Galactic center, $V$ in the direction of Galactic rotation, and $W$ vertically upwards out of the plane of the disk. We associate these three velocity coordinates with the $(\vR, \vtheta, \vz)$ coordinates of the simulation particles. Note that for consistency with the simulation, we set the azimuthal component of $\vec{v}_{\rm LSR}$ equal to the rotational velocity of the star particles in Eris ($205 \kms$), rather than to the IAU standard value of $220 \kms$.

The resulting Earth rest frame speed distributions are shown in Figure~\ref{fig:fv_earth_2panel_darkdisk} for $t=150.2$ days since J2000.0 (beginning of June), when the Earth's relative motion with respect to the Galactic DM halo is maximized.  The dark disk's azimuthal velocity distribution (left panel) peaks at $-85 \kms$ ($60 \kms$ below the peak of the stars). The additional low azimuthal speed material from the dark disk results in a marked excess in the low speed tail ($<200 \kms$) of the full speed distribution (right panel), and a corresponding deficit at intermediate speeds (200--400 \kms), in the Eris disk annulus compared to the ErisDark spherical shell. At even higher speeds ($>400 \kms$) we again see an excess in Eris compared to ErisDark. This high speed excess is the result of the overall broadening of the velocity distributions caused by the adiabatic contraction of the DM halo (the second process mentioned at the end of Sec.~\ref{sec:darkdisk}).

\subsection{Scattering Signal}

\begin{figure*}[htp]
\includegraphics[width=0.49\textwidth]{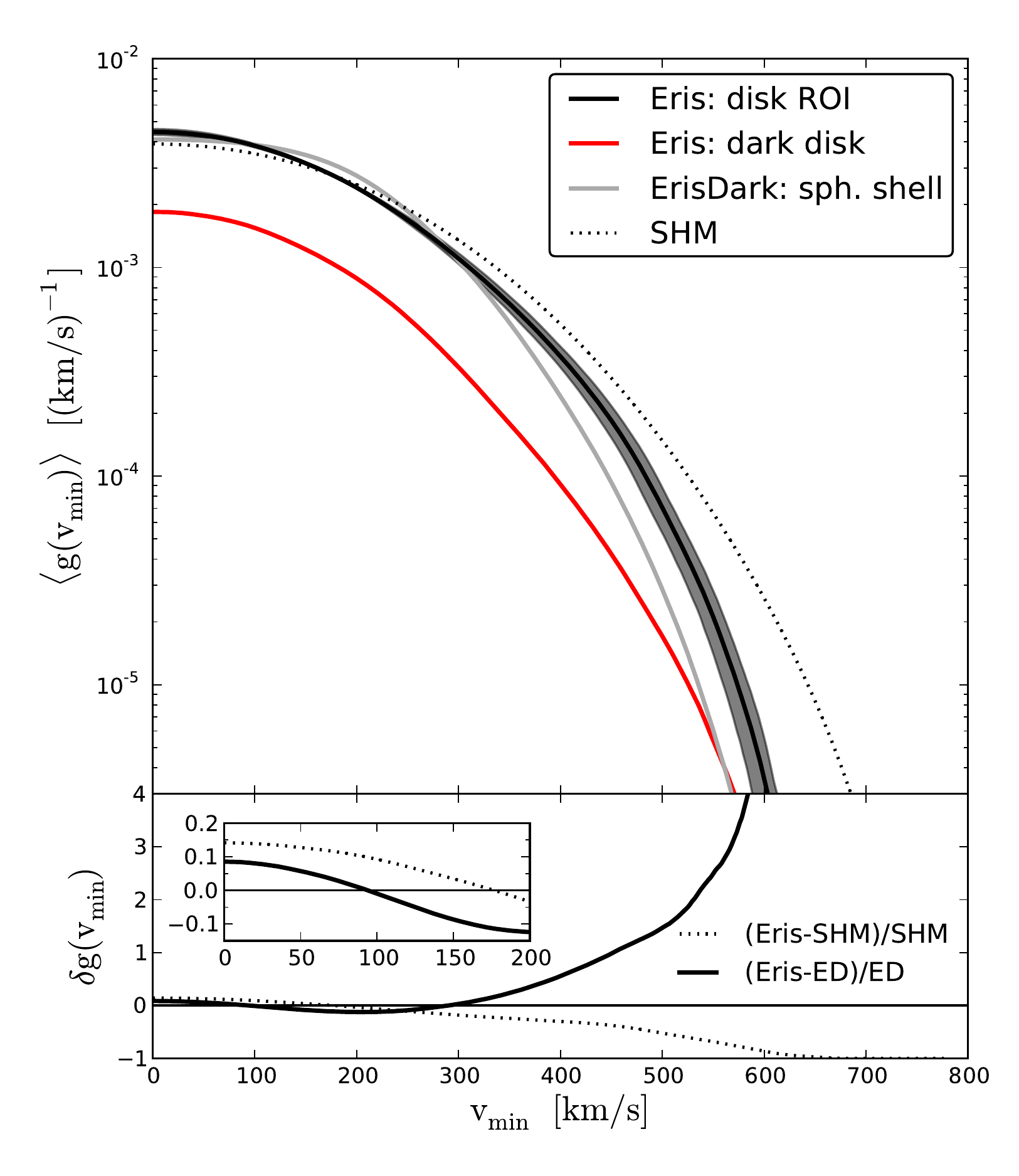}
\includegraphics[width=0.49\textwidth]{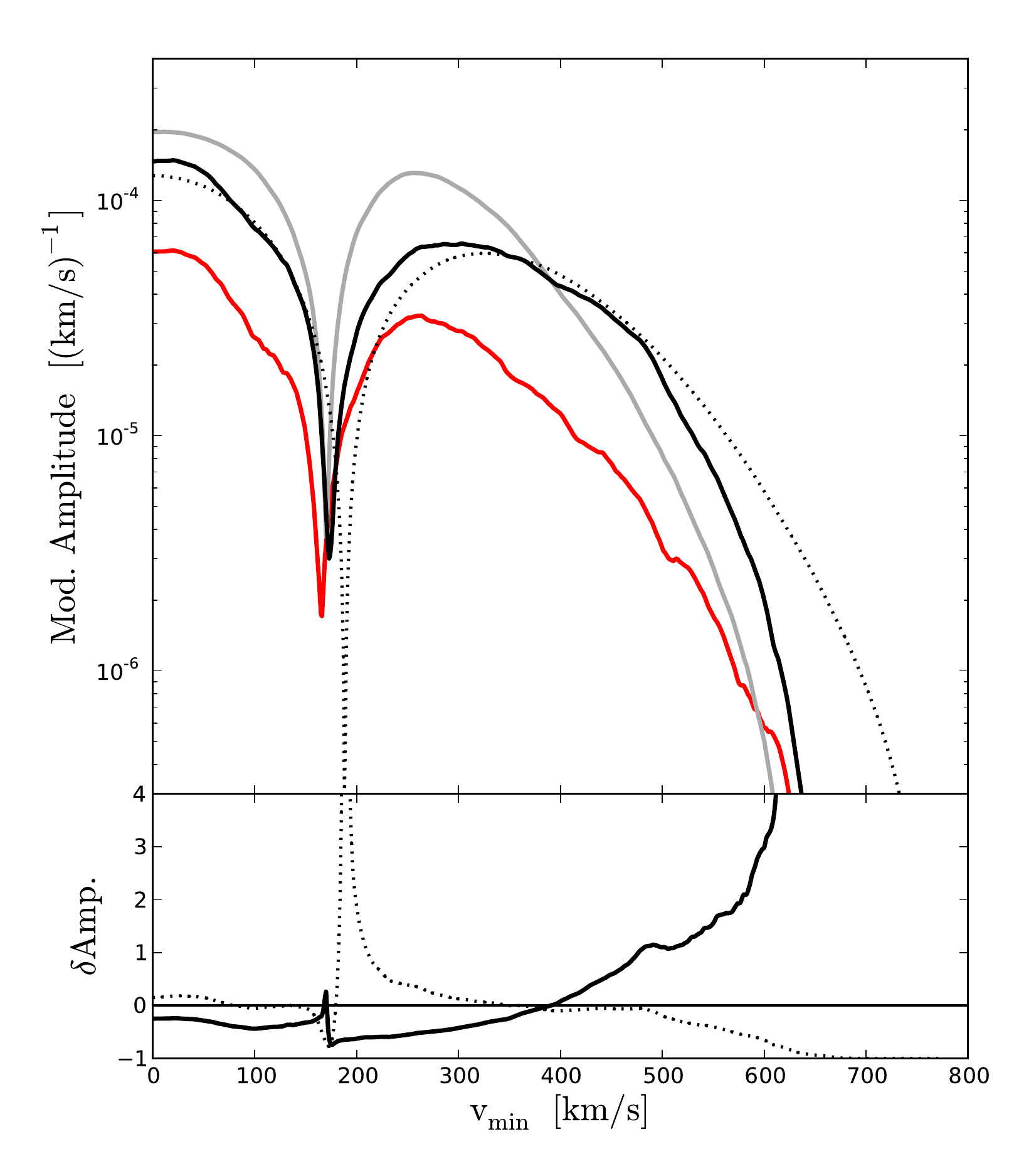}
\includegraphics[width=0.49\textwidth]{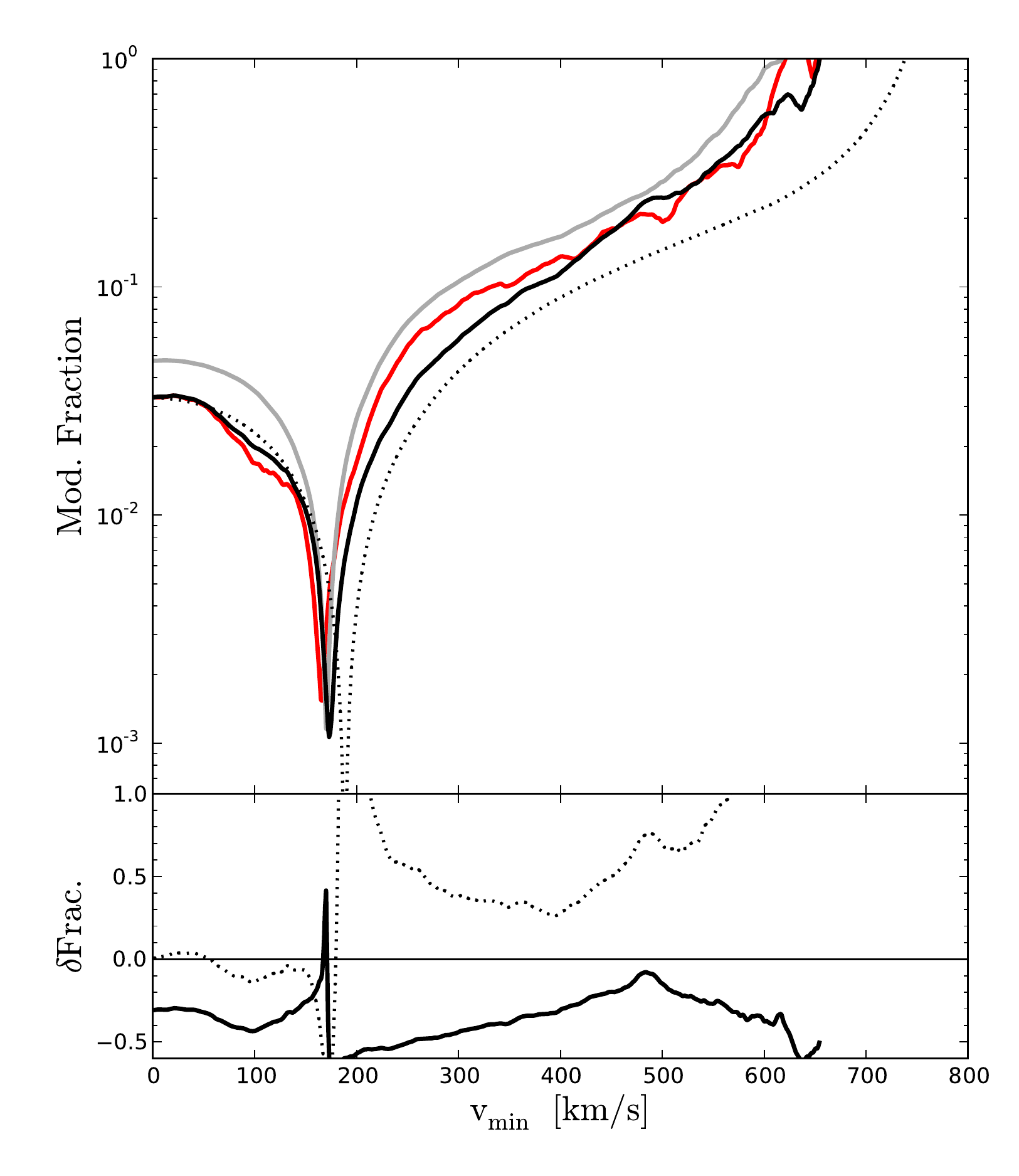}
\includegraphics[width=0.49\textwidth]{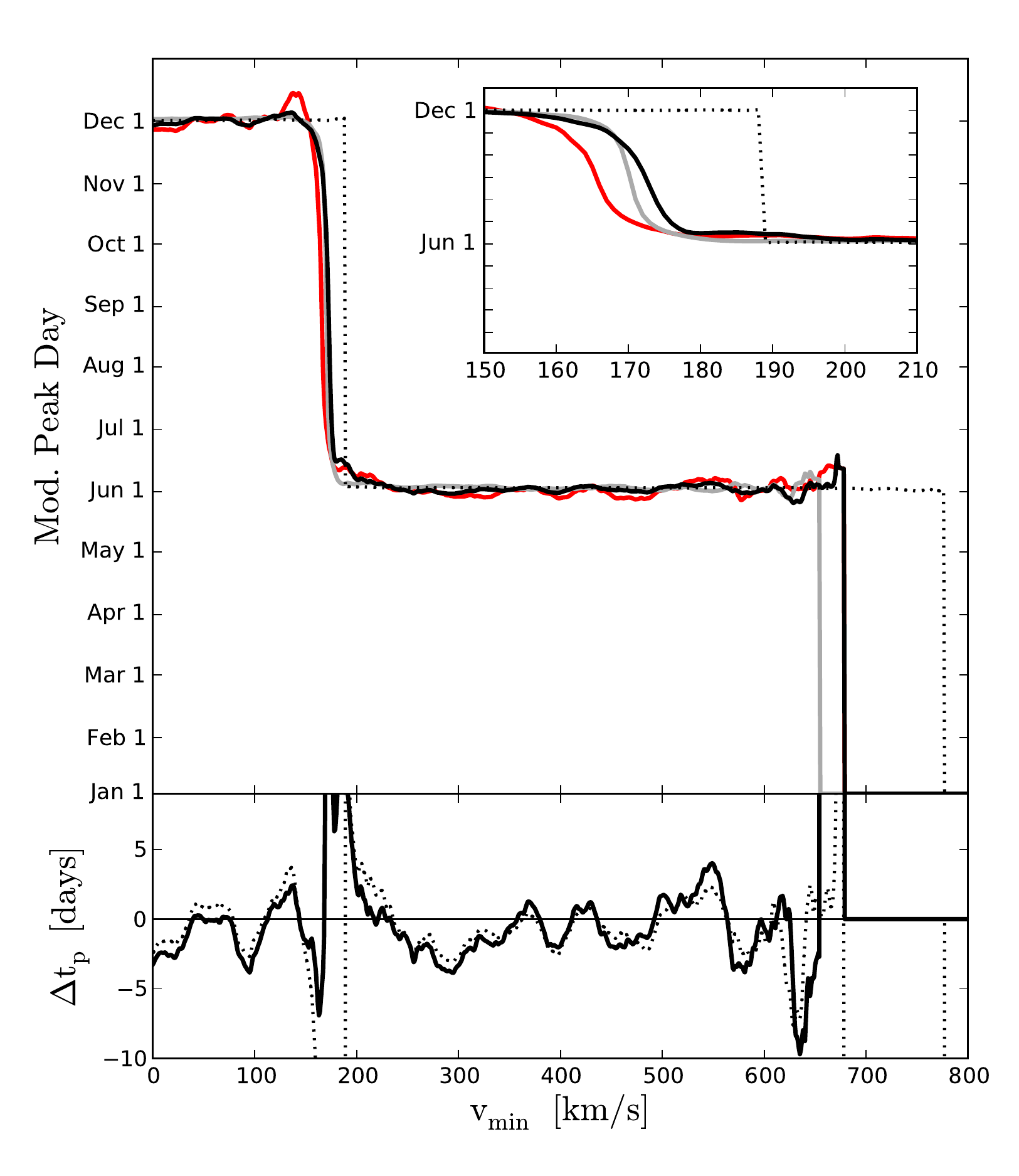}
\caption{Comparison of $g(v_{\rm min}) \equiv \int_{v_{\rm min}}^\infty f(v)/v \, dv$ and its annual modulation between Eris (black solid), the dark disk component in Eris (red solid), the ErisDark spherical shell (dashed), and Maxwellian (dotted). The scattering rate is directly proportional to $g(v_{\rm min})$, and we have fit it to a sinusoidal variation with a constant offset ($A + B \cos(2\pi (t-t_p)/365 {\rm d})$). The four panels show different components of this fit. In the bottom panels we show the fractional difference $\delta$ (absolute difference $\Delta$ for the peak day) between Eris and ErisDark (solid) and between Eris and the Maxwellian (dotted). \textit{Top left:} The time average (over one year), $A$. The shaded band covers the modulation amplitude ($A \pm B$). \textit{Top right:} The modulation amplitude, $B$. \textit{Bottom left:} The modulation fraction, $B/A$. \textit{Bottom right:} The peak day of the modulation, $t_p$. The inset shows a zoom-in of the region where the modulation phase flips.}
\label{fig:signal}
\end{figure*}

The differential DM-nucleus scattering rate per unit detector mass is given by \citep{lewin_review_1996}
\begin{equation}
\f{dR}{dE_r} = \f{\rho_0}{m_N \, m_\chi} \, \sigma(E_r) \, \int_{\vmin}^\infty \!\! \f{f(v)}{v} \, dv,
\end{equation}
where $\rho_0$ is the local (at Earth) DM density, $m_N$ is the mass of the target nucleus, $m_\chi$ the mass of the DM particle, $E_r$ is the recoil energy, $\sigma(E_r)$ is the energy-dependent scattering cross section, and $v = |\vec{v}|$ is the Earth-frame speed of the DM particles incident on the detector. When a DM particle with incident speed $v$ elastically scatters off a nucleus, it imparts some fraction of its kinetic energy as a nuclear recoil. A given $E_r$ can be produced by any particle with $v > \vmin = \sqrt{E_r \, m_N/(2\mu^2)}$ (where $\mu = m_N m_\chi / (m_N + m_\chi)$ is the reduced mass), and hence the differential scattering rate is directly proportional to
\begin{equation}
\gvmin \equiv \int_{\vmin}^\infty \!\! \f{f(v)}{v} \, dv.
\label{eq:gvmin}
\end{equation}

Not all DM direct detection experiments are sensitive to the same range of \vmin. For a given $m_N$ and an assumed $m_\chi$, the recoil energy sensitivity band $(E_{r,\rm min},E_{r,\rm max})$ of an experiment can be mapped onto a corresponding band in \vmin\ \citep{fox_integrating_2011,frandsen_resolving_2012}. Experiments with heavier nuclei and lower energy thresholds have smaller \vmin, and hence probe more of $f(v)$. In Table~\ref{tab:benchmarks} we list \vmin-bands for a number of prominent direct detection experiments and several representative values of $m_\chi$. Note that these combinations of experiments and DM particle masses cover almost the entire range of possible \vmin\ values, from $\vmin \lesssim 100 \kms$ for Xenon100 and CDMS-II(Ge) with a very massive DM particle ($m_\chi \sim 500$ GeV) all the way to beyond the escape speed for light DM particles ($m_\chi \lesssim 10$ GeV).

In Figure~\ref{fig:signal} we show comparisons of the average scattering rate $\langle \gvmin \rangle$ and the modulation amplitude, fraction, and peak day between Eris, the dark disk component of Eris, the ErisDark spherical shell, and the SHM model (Maxwellian with $\vpeak = 220 \kms$ and $v_{\rm esc} = 550 \kms$). Note that we compare to the SHM, rather than to the peak-matched Maxwellian with $\vpeak = 195 \kms$, since the SHM is commonly used in the direct detection literature. We have also chosen not to scale up the Eris velocities to match $\vpeak = 220 \kms$ (as was done in Kuhlen et al. 2010\nocite{kuhlen_dark_2010}, for example), since Eris after all is a realistic Milky Way analog galaxy, and so its lower $\vpeak$ should be considered a realistic possibility.

The top left panel of Figure~\ref{fig:signal} shows the time-averaged mean scattering rate, $\langle \gvmin \rangle$. The differences between Eris, ErisDark, and the SHM remain fairly modest at low and intermediate \vmin. Below 100 \kms\ the presence of the dark disk raises \gvmin\ by a few percent compared to ErisDark. Between 100 and 300 \kms, the dark disk results in a comparable reduction of the average scattering rate, but at even larger \vmin\ the non-rotating density enhancement again reverses the trend. The difference between Eris and ErisDark continues to grow monotonically, reaching a factor 2 (3, 4) enhancement near $\vmin=500$ (550, 600) \kms. Compared to the SHM, the average scattering rate in Eris is enhanced by up to 15 percent at $\vmin < 180 \kms$, but is reduced for all high \vmin. To summarize, the rotating dark disk components affects the average scattering rate only modestly, but the additional contraction (non-rotating density enhancement) strongly affects the scattering rates at large $\vmin$, leading to an increase compared to the DM-only simulation and a suppression compared to the SHM.

\subsection{Modulation}

Owing to the orbital motion of the Earth around the Sun, the speed with which particles impinge on Earth is modulated in the Earth's rest-frame on an annual time scale, and this modulation is propagated into $\gvmin$ \citep{drukier_detecting_1986}. We have fit the fully modulated \gvmin\ to a sinusoidal variation with a constant offset, $\gvmin(t) = A + B \cos(2 \pi (t - t_p)/365 {\rm d})$. The constant term ($A$) was discussed in the previous section; we now discuss the modulation amplitude ($B$, top right of Figure~\ref{fig:signal}), the modulation fraction ($B/A$, bottom left of Figure~\ref{fig:signal}), and the peak day ($t_p$, bottom right of Figure~\ref{fig:signal}).

The amplitude of the modulated component ($B$) decreases with increasing $\vmin$, which is simply a reflection of the decreasing average scattering rate. As discussed in more detail below, the phase of the modulation flips by 180 degrees at $\vmin \approx 175 \kms$, and this results in a null in the modulation amplitude. Note that even the dark disk component by itself (red line in Figure~\ref{fig:signal}) exhibits a small amount of modulation, including its own null. This is a result of the small lag between the dark disk and the stellar disk and the Sun's peculiar motion. If the dark disk were truly perfectly co-moving with the Sun, then Earth's orbital motion around the Sun would not result in any annual modulation. Of course in an actual experiment there is no way to distinguish whether a given scattering event is from a dark disk particle or from the background halo.

Compared to ErisDark, the presence of the dark disk suppresses the modulation by several tens of percent at $\vmin < 400 \kms$, except around 175 \kms, where the slight shift in the location of the modulation null leads to a small positive peak in the fractional difference. At $\vmin > 400 \kms$ the modulation amplitude in Eris begins to exceed that of ErisDark and again the difference quickly increases to factors of a few. Comparing to the SHM, the situation is reversed: below $400 \kms$ the modulation amplitude in Eris is mostly greater than in the SHM, but at higher speeds it drops below. Similar trends can be seen in the modulation fraction ($B/A$), except that here the Eris curve remains predominantly above the ErisDark and below the SHM curve for the entire range of $\vmin$. The inclusion of dissipational baryonic physics tends to decrease the modulation fraction at all speeds, but compared to the SHM the Eris (and ErisDark) simulation actually exhibits an enhanced modulation fraction.

The peak day of the modulation flips from occurring in the Northern summer (near June 1) at $\vmin \gtrsim 175 \kms$ to the winter (near December 1) at lower \vmin. This well-understood effect \citep{lewis_phase_2004,purcell_dark_2012} is a consequence of the shift of \fv\ to lower speeds from summer, when the relative motion between Earth and the DM halo is maximized, to winter, when it is minimized. Below some speed $f(v;{\rm winter})$ exceeds $f(v;{\rm summer})$, and vice versa at higher speeds. Since \gvmin\ is defined as an integral of $f(v)/v$ from \vmin\ to infinity (see Eq.~\ref{eq:gvmin}), this implies that there exists some speed, slightly below the peak of \fv\ ($0.89 \, v_{\rm peak}$ for a Maxwellian), for which $g(\vmin;{\rm winter}) = g(\vmin;{\rm summer})$. Below this speed the modulation peaks in the winter, above it in the summer, i.e. the phase of the annual modulation flips. The inset in the bottom right panel of Figure~\ref{fig:signal} shows that this transition is shifted by a few \kms\ to higher \vmin\ in Eris compared to ErisDark, but occurs at about 20 \kms\ less than in the SHM. Note also that the transition in the simulations is less abrupt than in the Maxwellian model.

The exact day on which the peak of the annual modulation occurs depends on the detailed shape of \fv, and it has been shown \citep{kuhlen_dark_2010,purcell_dark_2012} that DM velocity substructure can occasionally lead to marked changes (tens of days) in $t_p$, especially at large \vmin. The addition of baryonic physics, however, seems to have a more moderate effect on $t_p$. With an exception at the modulation null, $t_p$ does not change by more than three days between Eris and ErisDark and the SHM model.\\

\section{Summary}\label{sec:summary}

We have analyzed the local (6--10 kpc) DM distribution in Eris, one of the highest resolution N-body+SPH hydrodynamics simulations to date of the formation of a Milky-Way-like galaxy in a cosmological context. The simulated disk galaxy matches many observational constraints on the structure of the Milky Way, such as having an extended rotationally supported stellar disk, a gently falling rotation curve at 8 kpc, falling on the Tully-Fisher relation, having a stellar-to-total mass ratio of 0.04, a star formation rate of 1.1 M$_\odot$ yr$^{-1}$, a low bulge-to-disk ratio of 0.35, and even a hot halo with a pulsar dispersion measure in excellent agreement with measurements towards the Magellanic Cloud \citep[for more details, see][]{guedes_forming_2011}. Eris is the most realistic such simulation available today.

The focus of our study has been to assess the influence of dissipational baryonic physics on the DM density and velocity distribution at the location of the Sun, and its implications for Earth-bound DM direct detection experiments. To this end we have also analyzed the ErisDark simulation, a DM-only counterpart to Eris, using the same initial conditions except that all matter is treated as collisionless DM. Direct comparisons between Eris and ErisDark allow us to isolate the effects of the baryonic physics. We have also compared Eris to the Standard Halo Model (Maxwellian with $\vpeak = 220 \kms$, $v_{\rm esc} = 550 \kms$), in order to highlight changes relative to this simplified model, which is still commonly used in the direct detection literature.

The main results of our study are summarized as follows:
\begin{itemize}

\item The local DM density at 8 kpc in the disk plane in Eris is 0.42 GeV cm$^{-3}$, about 34\% higher than the Eris spherical average and 31\% higher than the ErisDark spherical average. This indicates that the dissipational baryonic physics in Eris has led to a contraction of the dark matter halo, and that this contraction is most pronounced in the disk plane.

\item In our disk region-of-interest (ROI, an annulus centered at 8 kpc with width and height equal to 4 kpc) the distributions of radial and azimuthal velocity components are broadened in Eris compared to ErisDark, and only slightly narrower in the vertical component. As a result, the speed (velocity modulus) distribution is also broadened and shifted to higher speeds. This reflects the deeper potential well created by the dissipation of the baryons, which have sunk to the center of the halo. Nevertheless, the peak of the speed distribution in Eris occurs at only $\vpeak = 195 \kms$, considerably below that of the SHM ($\vpeak = 220 \kms$).

\item As observed in DM-only simulations, the speed distribution in Eris is not perfectly described by a Maxwellian shape, exhibiting a deficit at speeds below its peak and an excess at higher speeds. However, the differences to the peak-matched Maxwellian are much smaller in Eris than in ErisDark.

\item Both the ErisDark and Eris \fv\ are well described by the empirical fitting function recently proposed by \citet{mao_halo--halo_2013}. The best-fit value of $p$ (a parameter governing how steeply the high speed tail falls) is higher in Eris (2.7) than in ErisDark (1.5). A more steeply falling \fv\ eases the tension between non-detections in direct detection experiments with heavy nuclei (e.g. Xenon-100) and tentative signals from experiments with lighter nuclei (e.g. CDMS-Si, CoGeNT).

\item The azimuthal velocity component in Eris (but not in ErisDark) is noticeably skewed towards positive \vtheta, with 30\% more prograde than retrograde (with respect to the stars) material. This indicates the possible presence of a ``dark disk''. 

\item We have quantified the Eris dark disk component by following the accretion history of the 160 most massive satellites. 81\% of all accreted material in the disk ROI comes from satellites with positive asymmetry parameter Asym, i.e. depositing more prograde than retrograde rotating material. We define as a ``dark disk'' all material deposited by satellites with high asymmetry, $\Asym > 2/3$. With this definition, the dark disk in Eris contributes 9.1\% (0.034 GeV cm$^{-3}$) of the DM density in the disk ROI. Additionally applying the commonly used criterion that dark disk material lie within 50 \kms\ of the stellar rotation speed, the dark disk contribution drops to 3.2\% (0.012 GeV cm$^{-3}$).

\item The dark disk contributes only about one third of the excess DM density in the Eris disk ROI over the ErisDark spherical average (0.12 GeV cm$^{-3}$), and this suggests that there are at least two distinct processes leading to an enhancement of the DM in the disk plane: one process that results in a DM component with significant net angular momentum and that is nearly co-rotating with the stellar disk, and another process that pulls DM into the disk plane without forcing it to co-rotate \citep[see also][]{zemp_impact_2012}.

\item The time-averaged scattering rate, proportional to \gvmin, exhibits only mild changes from ErisDark to Eris for most values of \vmin. At very low \vmin\ \mbox{($< 100 \kms$)}, the co-rotating dark disk component leads to a few percent increase in \gvmin, since there are slightly more particles with low relative speeds. Bigger changes are seen at high \vmin, where the broadening of \fv\ due to the overall halo contraction leads to scattering rates that are several times higher than in ErisDark. On the other hand, comparing to the SHM, the mean scattering rate is strongly reduced at high \vmin.

\item Similar trends hold for the amplitude of the annual modulation in Eris. Compared to ErisDark, it is slightly suppressed at low \vmin, and strongly enhanced at high \vmin. Compared to the SHM, however, the modulation amplitude is suppressed, just like the non-modulating part. The sign of the effect is reversed for the modulation fraction: it is suppressed by $\sim 50\%$ with respect to ErisDark, but similarly enhanced compared to the SHM, across the whole range of \vmin. Lastly, the peak day of the modulation is not strongly affected by the dissipational physics, with changes typically not exceeding $\pm 3$ days at most \vmin. Compared to the SHM, however, the \vmin\ corresponding to the sign flip in the modulation phase shifts by about 15 \kms. \\

\end{itemize}

In conclusion, we have in this work investigated the effects dissipational baryonic physics has on the local distribution of DM near the Sun. We are able to isolate these effects through a comparative analysis of two twin cosmological galaxy formation simulations with identical initial conditions, one of which (Eris) being a full hydrodynamic simulation and the other (ErisDark) a DM-only one. Since the Eris simulation results in a realistic Milky Way analog galaxy, its DM halo can be viewed as a more realistic alternative to the Maxwellian standard halo model commonly used in analysis of direct detection experiments.

As DM direct detection experiments continue to develop and become ever more sensitive, it will be of paramount importance to properly understand and quantify the expectations provided by realistic simulations of galaxy formation. We look forward to the day when large numbers of detected DM scattering events will allow direct tests of these predictions.

\section*{Acknowledgments}

Support for this work was provided by the NSF through grant OIA-1124453, and by NASA through grant NNX12AF87G (P.M.).  
J.G. was funded by the ETH Zurich Postdoctoral Fellowship and the Marie Curie Actions for People COFUND Program. The Eris Simulation was carried out at NASA's Pleiades 
supercomputer, the UCSC Pleiades cluster, and ErisDark was performed at the UCSC Pleiades cluster.

\bibliographystyle{apj}
\bibliography{ErisDarkDisk}

\end{document}